\documentclass[11pt]{article}

\usepackage{acl}

\usepackage{times}
\usepackage{latexsym}
\usepackage[T1]{fontenc}
\usepackage[utf8]{inputenc}
\usepackage{microtype}
\usepackage{inconsolata}
\usepackage{graphicx}

\usepackage{amsmath,amssymb,bm}
\usepackage{booktabs}
\usepackage{enumitem}
\usepackage{tabularx}
\usepackage{array}
\usepackage{multirow}
\usepackage{tikz}
\usepackage{xcolor}
\usepackage{etaremune}
\usetikzlibrary{arrows.meta,calc,positioning,shapes.geometric}

\newcommand{\overviewfallback}{
\begin{tikzpicture}[
    x=1cm,y=1cm,>=Latex,
    box/.style={draw, rounded corners, align=center,
                minimum width=2.7cm, minimum height=1.25cm, inner sep=4pt,
                font=\footnotesize},
    semnode/.style={draw, dashed, rounded corners, align=center,
                    minimum width=3.2cm, minimum height=0.9cm, inner sep=3pt,
                    font=\scriptsize},
    evalbox/.style={draw, rounded corners, align=center,
                    minimum width=6.8cm, minimum height=1.1cm, inner sep=4pt,
                    font=\footnotesize}
]
    \node[box] (detect) at (0,0) {(1) Skill detectors\\empathy, exploration};
    \node[box] (state)  at (3.6,0) {(2) State accumulation\\score $D_t$};
    \node[box] (disc)   at (7.2,0) {(3) Threshold\\3 levels (G/M/H)};
    \node[box] (llm)    at (10.8,0) {(4) LLM generator\\tiered card + context};
    \node[semnode] (sem) at (3.6,1.8) {SEM-implied ratio\\(prior work)};
    \node[font=\scriptsize, align=center] at (-1.5,0) {Trainee\\turn};
    \node[font=\scriptsize, align=center] at (12.6,0) {Patient\\response};
    \node[evalbox] (eval) at (5.4,-2.0)
        {Evaluation: three adaptivity properties + human ratings};
    \draw[->,thick] (-0.9,0) -- (detect);
    \draw[->,thick] (detect) -- (state);
    \draw[->,thick] (state) -- (disc);
    \draw[->,thick] (disc) -- (llm);
    \draw[->,thick] (llm) -- (12.0,0);
    \draw[->,thick,dashed] (sem) -- (state);
    \draw[->,thick,dotted] (state.south) -- (eval.north);
    \draw[->,thick,dotted] (llm.south) |- (eval.east);
\end{tikzpicture}
}

\newcommand{\uifallback}{
\begin{tikzpicture}[x=1cm,y=1cm,>=Latex]
    \draw[rounded corners] (0,0) rectangle (14,6);
    \draw (3.6,0) -- (3.6,6);
    \draw (0,3.1) -- (3.6,3.1);
    \draw (3.6,0.9) -- (14,0.9);
    \node[font=\small\bfseries] at (1.8,5.45) {VP Selection};
    \node[font=\scriptsize,align=left] at (1.8,4.45) {Sam1\\Sam2\\Alex1\\Alex2};
    \node[font=\small\bfseries] at (1.8,2.55) {Patient Info};
    \node[font=\scriptsize,align=left] at (1.8,1.65) {Persona summary\\background\\presenting concerns};
    \node[font=\small\bfseries] at (8.8,5.45) {Chat Workspace};
    \draw[rounded corners] (4.1,1.25) rectangle (13.5,5.0);
    \node[font=\scriptsize,align=left] at (5.9,4.3) {Therapist: ...};
    \node[font=\scriptsize,align=left] at (10.9,3.5) {Patient: ...};
    \node[font=\scriptsize,align=left] at (5.9,2.7) {Therapist: ...};
    \node[font=\scriptsize,align=left] at (10.9,1.9) {Patient: ...};
    \node[font=\small\bfseries] at (8.8,0.45) {Input / Restart Controls};
\end{tikzpicture}
}

\title{Empirically Grounded Behavioral Adaptivity for LLM-Based Virtual Patients in Psychotherapy Training}

\author{%
Angela H. Chen$^{1}$%
~\;~\;~Siwei Jin$^{2,*}$
~\;~\;~Catherine Bao$^{3,*}$
~\;~\;~Canwen Wang$^{1}$
\\
\textbf{Yi-Chia Wang$^{1}$
~\;~\;~Robert Kraut$^{1}$
~\;~\;~Tongshuang Wu$^{1}$
~\;~\;~Haiyi Zhu$^{1,\dagger}$}\vspace{0.5ex}
\\
$^{1}$Carnegie Mellon University
~\;~\;~
$^{2}$Independent Researcher
~\;~\;~$^{3}$University of Utah
\\
\texttt{\{angelachen, haiyiz\}@cmu.edu}
}

\begin{document}
\maketitle

\renewcommand{\thefootnote}{\fnsymbol{footnote}}
\footnotetext[1]{Equal contribution; work done at CMU.}
\footnotetext[2]{Corresponding author.}
\renewcommand{\thefootnote}{\arabic{footnote}}
\setcounter{footnote}{0}

\renewenvironment{quote}
  {\list{}{\leftmargin=0.4cm \rightmargin=0.4cm}%
   \item\relax\looseness=-1}
{\endlist}

\begin{abstract}
Simulated patients offer a scalable way to train psychotherapy micro-skills such as empathic responding and exploratory probing, but current systems either follow fixed scripts or rely on LLMs that drift unpredictably over long sessions. We present the Adaptive Virtual Patient (AVP), which adapts its disclosure behavior in response to trainee skill, from guarded, through moderate openness, to full disclosure. The AVP is informed by a structural equation model fit to nearly 2{,}000 hours of real-world psychotherapy transcripts, which quantifies how therapist empathy and exploration shift a patient's openness over time. An LLM generates the AVP's utterances conditioned on a disclosure level that the dynamics module updates each turn. In an evaluation with 20 clinicians and trainees over 80 sessions (1{,}033 turns), the AVP's disclosure rises in response to therapist empathy and exploration, while a prompt-only baseline stays flat; ablations confirm that the empirically motivated parameterization outperforms alternatives, with exploration carrying most of the adaptive signal.

\end{abstract}


\begin{figure}[h!]
  \centering
  \IfFileExists{tease6.pdf}%
    {\includegraphics[width=\linewidth]{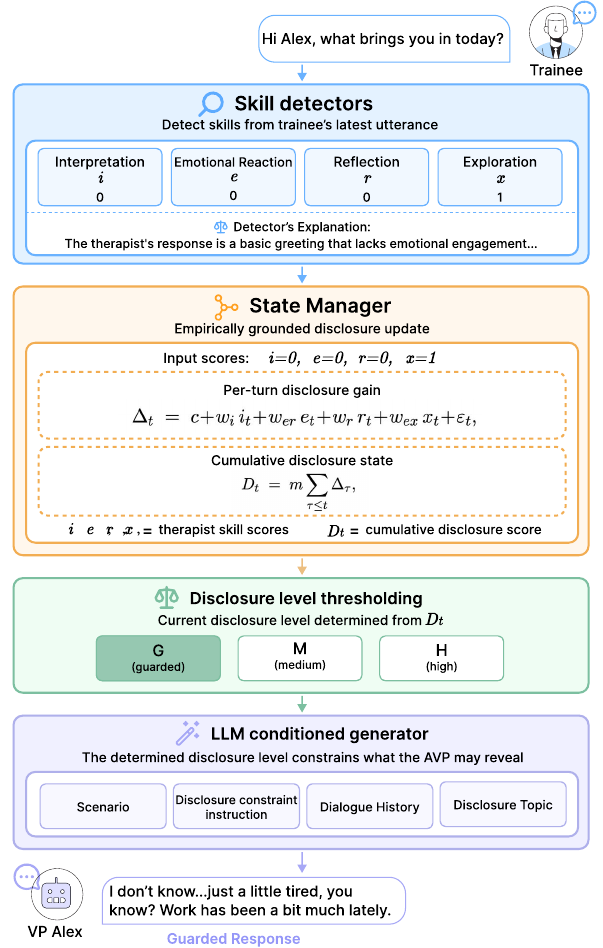}}%
    {\overviewfallback}
  \caption{Overview of our Adaptive Virtual Patient (AVP) framework, where we explicitly control the VP's behavioral state via an explicit intermediate computation, such that the VP's openness directly ties with the trainee performance.
 }
  \label{fig:overview}
\end{figure}

\section{Introduction}\label{sec:introduction}

Large language models (LLMs) are increasingly used as conversational agents in education, training, and social-interaction research~\citep{shao-etal-2023-character,
martynova-etal-2025-llms, park2023generative}. They produce fluent and plausible responses, but their long-horizon behavior is governed by a static persona prompt, which limits the within-session change~\citep{wang-etal-2024-rolellm, samuel-etal-2025-personagym}. An agent may sound like a patient, student, or customer on each turn but fail to adapt in ways that reflect an evolving state in relation to what the user does~\citep{abdulhai2026consistently, li2024measuringcontrollinginstructioninstability}. We study this problem in psychotherapy training, where agents simulate patients
with different mental health conditions~\citep{kenny2007virtual, tanana2019development}.

A realistic session with a guarded patient unfolds like this: if the trainee asks closed, agenda-driven questions, the patient gives brief, surface-level answers; if the trainee follows up with open-ended probes (``what was that like for you?'') and empathic reflections (``that must have been hard''), the patient begins to share more personal, emotionally loaded details. This contingent response is itself feedback for the trainee; without it, the trainee cannot learn what works~\citep{doi:10.3102/003465430298487, article}.

Despite the promise of using simulated virtual patients (VPs) to scale up the psychotherapy training, existing simulation approaches only partially support this adaptive cycle. Scripted VPs are reproducible but their predefined branches limit turn-by-turn adaptation~\citep{woodham2015medical, info:doi/10.2196/14676}; While prompt-only LLM patients support open, fluent exchanges~\citep{lozoya2025leveraging, wang-etal-2024-rolellm}, their behavioral change emerges implicitly from the prompt rather than from the trainee's input. Recent adaptive-VP work modifies patient behavior using trainee evaluation scores~\citep{lee2025adaptivevpframeworkllmbasedvirtual}, but its rule-based mapping from scores to behavior is not grounded in psychotherapy theory or data and thus does not model the gradual progression by which therapist behaviors shape a patient's state over time.

In response, we propose an \textbf{Adaptive VP} framework that separates \emph{behavioral state transitions} from \emph{language generation}. As shown in Figure~\ref{fig:overview}, after each trainee turn, LLM evaluators score clinically meaningful behaviors (empathy and exploration); a dynamics layer accumulates these scores into an ordinal disclosure level (low, medium, high); and an LLM generates the VP's next utterance conditioned on that level. Making the behavioral state an explicit intermediate computation keeps the VP's adaptation controllable and testable.

Specifically, we design the core dynamics layer to be informed by a structural equation model (SEM) estimated from over 2{,}000 hours of human-to-human psychotherapy session transcripts, which gives directed effects of therapist behaviors on next-turn patient disclosure~\citep{chen2026therapist_client_dynamics}. The deployed system uses integer weights approximating the SEM-implied empathy-to-exploration ratio. Self-disclosure is the controlled behavioral construct: the deployed system's discrete disclosure level can be checked against an independent rater's coding turn by turn.

We evaluate the framework in a within-subjects experiment: 20 clinicians and trainees held 80 brief conversations (1{,}033 turns) with four virtual patients varying in persona and system type. We test whether trainee behaviors produce coherent disclosure trajectories, whether generated utterances align with the intended disclosure level, and whether the patient is perceived as realistic, adaptive, and useful for training. With the adaptive virtual patient, disclosure climbs coherently across the session and matches the system's internal state in the generated text --- neither of which the prompt-only baseline achieves.

\paragraph{Contributions.} We make three contributions.
First, we introduce a virtual-patient framework that handles \textbf{behavioral change} as a small, explicit module between the trainee's turn and the VP's reply: it updates a latent disclosure state using weights derived from a psychotherapy-data structural equation model, and the LLM generates the next utterance conditioned on that state. Keeping the behavioral logic separate from generation makes the VP's adaptation inspectable, controllable, and fixed across experimental conditions in a way a prompt-only system cannot achieve. 

Second, we propose \textbf{four measurable properties} an AVP should satisfy: \emph{P1
Responsiveness} (the state responds to user behavior in the
direction the data predicts), \emph{P2 Gradual Build-up} (the state climbs steadily across the session and only as the user earns it), \emph{P3 Faithful Realization} (the generated text matches the system's intended state), and \emph{P4 Skill Sensitivity} 
(whether disclosure varies with user skill);
each property comes with its own metric, so we can locate exactly
where a system fails rather than collapse the question into one
number. 

Third, we provide \textbf{empirical evidence} from 20 clinicians and trainees, that the adaptive patient can produce a steadily climbing disclosure trajectory while a prompt-only baseline with the same LLM and persona stays flat, which better support the loop needed for psychotherapy training. 
On post-session ratings, the adaptive version of an expressive persona outscored its static counterpart on character consistency, adaptivity, realism, and training usefulness, with the strongest endorsements from licensed therapists; and a parameter ablation further shows the deployed weighting beats every alternative tested in agreement with independent ratings.\looseness=-1

\section{Background and Related Work}\label{sec:background}

\paragraph{Virtual patients and LLM-based role-play agents.}
Virtual and simulated patients have long been used as a mechanism help trainees 
practice clinical communication skills such as empathic
responding, reflective listening, and exploratory questioning
\citep{barrows1993standardized,kenny2007virtual,tanana2019development}.
Reviews and meta-analyses show that virtual patients can support
health-professions training at scale
\citep{cook2009virtualpatients,cook2011simulationmeta,consorti2012virtualpatients,kononowicz2019virtualpatients}.
Traditional systems often prioritize structured case progression and
reproducibility, while recent LLM-based systems enable more fluent
and open-ended patient simulation through prompt-based personas,
theory--informed cognitive models, or
domain-expert authored role-plays
\citep{wang2024patientpsi,holderried2024simulated, louie-etal-2024-roleplay,lozoya2025leveraging}.
More general generative-agent architectures further show that memory,
reflection, and planning can support believable long-horizon behavior
\citep{park2023generative}. Across these approaches, however,
behavioral change is typically governed by scripts, prompts, memory,
or qualitative principles rather than empirically estimated
interaction dynamics. Our work instead updates a latent disclosure
state from detected trainee behaviors and conditions language generation accordingly.\looseness=-1

\paragraph{Controllable dialogue generation.}
Prior work in controllable dialogue generation has shown that external
conditioning signals (e.g., personas, control codes, styles,
attributes, and latent discourse variables) can steer generated
responses
\citep{zhang2018personalizing,keskar2019ctrl,zhao2017learning,serban2017hierarchical}.
These methods generate language conditioned on a desired persona,
style, or latent variable, but generally do not specify how that
control variable should update as a function of an interlocutor's
behavior. Recent role-playing benchmarks similarly evaluate whether
LLMs can maintain assigned roles or personas over multi-turn
interactions
\citep{wang-etal-2024-rolellm,samuel-etal-2025-personagym}.  In
contrast, our setting requires behavior-contingent control: the target
state (i.e., the VP's willingness to self-disclose) itself must
evolve over time, and those changes should be inspectable and grounded
in an empirical model of interaction dynamics. Our framework provides
this control through an explicit state-transition step: detected
user behaviors update a latent disclosure state before the next
VP utterance is generated.

\paragraph{State-based dialogue management.}
State-based dialogue systems provide a precedent for separating state
estimation from response or action generation. In task-oriented
dialogue, dialogue state tracking (DST) estimates the values of
conversation-relevant slots (e.g., a restaurant's cuisine, price
range, or location), user goals, or belief states from conversation
history, and downstream policies use those states to select system
actions \citep{6407655, balaraman2021dst, jacqmin-etal-2022-follow}.
This separation makes dialogue behavior more controllable and
diagnosable than direct response generation alone. Our work adapts
this state-based view to simulated interpersonal behavior; the state
is not a task slot or user goal, but a continuous latent behavioral
variable, VP disclosure level, whose transition is driven by detected
user behaviors and parameterized by empirical psychotherapy data.
The next VP utterance is then generated conditioned on this
updated state.


\section{Method}\label{sec:methods}

\subsection{Empirically Grounded Adaptivity}\label{sec:adaptivity_definition}

We use ``empirically grounded adaptivity'' to mean four things
across a session (elaborated and quantified in \S\ref{sec:properties}): (i) the VP's internal state changes from
turn to turn in response to user behaviors, in a direction
consistent with an external data-driven model of interaction
dynamics; (ii) the state trajectory climbs gradually, advancing only as user competence accumulates; (iii) generated utterances
reflect the internal state rather than running independently of
it; and (iv) different user strategies produce different trajectories.

\paragraph{Constructs and Their Measurement}\label{sec:constructs}

To operationalize the adaptivity, our framework adds one explicit step between user input and VP output: an estimate of how the VP's willingness to self-disclose should change. 

More specifically, in our system the \textbf{user behaviors} are empathy and exploration scores: 
\begin{itemize}[leftmargin=1.2em,labelwidth=*,align=left, nosep]
    \item \emph{Empathy}: User behaviors communicating
        understanding of the VP, scored on a $0$-$2$ scale
        along three components: interpretation ($i$) and
        emotional reaction ($e$) from
        EPITOME~\citep{sharma2020computational}, and reflection
        ($r$) from MI~\citep{millerrollnick2023motivational}.
    \item \emph{Exploration}: User behaviors that invite the
        VP to elaborate, such as open-ended questions and
        probes; scored ($x$) on the same scale.
\end{itemize}

The \textbf{internal state} is an ordinal disclosure level derived from a cumulative score: 
\begin{itemize}[leftmargin=1.2em,labelwidth=*,align=left, nosep]
    \item \emph{Self-disclosure}: The VP's willingness to
        share personal or sensitive material on a given turn:
        external or factual chat (\textbf{G}uarded), specific personal content (\textbf{M}edium), or sensitive material such as suicidal ideation or named trauma (\textbf{H}igh). \looseness=-1
\end{itemize}

Finally, the \textbf{external model} is the structural equation model (SEM) of \citet{chen2026therapist_client_dynamics}. 

We make self-disclosure the controlled state for three reasons: It is a
clinically central indicator of therapeutic progress that empathy and
exploration are meant to elicit; It is the dependent variable the underlying SEM predicts; so the empirical grounding transfers to the update rule; And, it is observable turn by turn from the transcript,
which is what makes it usable as a controllable, evaluable state.
Disclosure is one dimension of patient behavior; extending the state
to affect and therapeutic alliance is deferred to future work.

\subsection{System Overview}\label{sec:method_overview}

As in Figure~\ref{fig:overview}, our framework explicate adaptivity by running  three components at each turn:

\begin{enumerate}[leftmargin=1.2em,labelwidth=*,align=left, nosep]
    \item \textbf{Skill detection.} Following \citet{chen2026therapist_client_dynamics}, we use two LLM evaluators to score the user's most recent turn on three \emph{empathy} components (interpretation, emotional reaction, reflection) and on \emph{exploration}, each on a $0$-$2$ scale.
    \item \textbf{State update.} We build a dynamics state management  module (\S\ref{subsec:state-manage}), which adds the scored turn to a running cumulative score $D_t$, with weights approximating the empathy-to-exploration ratio estimated from psychotherapy transcripts~\citep{chen2026therapist_client_dynamics}. $D_t$ is binned into one of three ordinal disclosure levels (\textbf{G}uard, \textbf{M}edium, \textbf{H}igh) by fixed thresholds (Table~\ref{tab:disclosure_thresholds}).
    \item \textbf{Conditioned response} (\S\ref{sec:generation}). We use an LLM to write the VP's next utterance from four prompt blocks: a disclosure-constraint instruction indexed by the current level, a level-matched disclosure topic block, the scenario brief, and recent dialogue history. The level determines \emph{what} the patient may reveal; the LLM determines \emph{how} it's said.\looseness=-1
\end{enumerate}


\paragraph{Models and detector validation}\label{sec:models}
We use \texttt{gpt-4o-mini} throughout the pipeline for three roles: (i)
generating the VP's responses, (ii) scoring the
user's empathy and exploration on each turn (the skill detectors that drive the state update), and (iii) assigning a post-hoc disclosure label (G/M/H) for each VP utterance, used as the ground-truth signal in the property analyses
(\S\ref{sec:desiderata_results}). All LLM calls share the scoring approach of
\citet{chen2026therapist_client_dynamics}, who validated this
model against expert annotators on a held-out human-coded subset of their corpus; all five reached good agreement with trained annotators (ICC(2,$k$) $\geq 0.60$; per-construct numbers and prompt details in Appendix~\ref{app:prompts}).\looseness=-1

\begin{table}[t]
\centering
\caption{Mapping from cumulative score $D_t$ to discrete disclosure levels.}
\label{tab:disclosure_thresholds}
\small
\begin{tabularx}{\columnwidth}{@{}l|l|X@{}}

\toprule
\textbf{$D_t$} & \textbf{Level} & \textbf{Description} \\
\midrule
$<4.5$        & \textbf{G}uarded & Surface-level; no personal details \\
$[4.5,\,10)$  & \textbf{M}edium Dis. & Specific personal content; not crisis \\
$\ge 10$      & \textbf{H}igh Dis. & Sensitive, emotionally significant material \\
\bottomrule
\end{tabularx}
\end{table}

\subsection{State management}\label{subsec:state-manage}

Intuitively, the state manager treats disclosure as a \emph{accumulating quantity}: each user turn adds an empathy-weighted and exploration-weighted increment to a running score, with small per-turn noise. Once accumulated, disclosure progress persists, and the score does not decay. However, the discrete label only advances when the score crosses a fixed threshold.\looseness=-1

At each user turn $t$, the \textbf{per-turn increment} is
\begin{equation*}
\Delta_t \;=\; c + w_i\,i_t + w_{er}\,e_t + w_r\,r_t + w_{ex}\,x_t + \varepsilon_t,
\end{equation*}
where $i_t, e_t, r_t \in \{0,1,2\}$ are the interpretation, emotional
reaction, and reflection scores; $x_t \in \{0,1,2\}$ is the
exploration score;
$\mathbf{w} = (w_i, w_{er}, w_r, w_{ex}) = (1, 1, 1, 3)$ are fixed
integer weights that approximate the SEM-implied
empathy-to-exploration ratio; $c = 0.15$ is a small positive constant
ensuring monotone progress under neutral input; and
$\varepsilon_t \sim \mathcal{N}(0, \sigma^2)$ with $\sigma = 0.10$
captures stochastic variation.

Conversely, the \textbf{cumulative score} is
\begin{equation*}
D_t \;=\; m \sum_{\tau \le t} \Delta_\tau,
\end{equation*}
where $m = 0.20$ is a global scaling factor (a ``difficulty
multiplier'' chosen to tune the typical session pace; lower values
delay arrival at H). The discrete disclosure label $L_t$ is resolved
by fixed thresholds (Table~\ref{tab:disclosure_thresholds}). We discretize the continuous score to enable explicit content gating and turn-by-turn evaluation against the same G/M/H coding scheme.

$D_t$
is monotone non-decreasing in expectation. What prevents premature
high-disclosure is the label thresholds (Table~\ref{tab:disclosure_thresholds}).
Under neutral user input ($i_t, e_t, r_t, x_t = 0$), each turn
adds only $m \cdot c = 0.03$ to the score, so reaching the M
threshold (4.5) takes roughly 150 turns of neutral input but as few
as 6 to 8 turns of high-skill input. This is the mechanism behind the
\emph{conditional} restraint we report in
\S\ref{sec:desiderata_results}: the AVP stays at G until the
user's accumulated empathy and exploration reach the next level.

\subsection{Response Generation}\label{sec:generation}

The level determines \emph{what} the VP may reveal; the LLM determines \emph{how} they say it. We do so by gating the disclosure topics both in the prompt and the memory retrieval. For example,  At G, H-level facts are not in the prompt so the LLM cannot produce them; long-term memory is also seeded only from G-level content at session start, so memory retrieval cannot surface sensitive material before the level allows it.

Specifically,  the disclosure level is resolved, the system builds the prompt from three parts: (1) a \emph{disclosure-constraint instruction} telling the LLM how openly the VP should behave at this level (Sam's G-level instruction reads: ``Sam does not share any personal information. She sticks to neutral or external topics, gives minimal responses, and stays vague but polite''); (2) a set of \emph{disclosure topics} drawn from one of three level-indexed sets per VP, structured around the graded intimacy framework for self-disclosure~\citep{bak2014selfdisclosure,balani2015detecting} (at G only neutral topics appear, e.g., ``work is stressful''; at H sensitive topics appear, e.g., a family member's suicide, dark thoughts); and (3) \emph{recent dialogue history} for local context. The VP's persona, voice, personality, and scenario remain fixed across levels. Full prompts and disclosure topics are in Appendix~\ref{app:personas}.

\paragraph{Example.} Sam is a teacher in her mid-30s with chronic work stress. Early in a session, while the cumulative score is still below the M threshold, her G-level prompt provides only neutral disclosure topics and an instruction to keep responses short and vague:\looseness=-1

\begin{quote}
\textbf{Trainee:} How long have you been at this job?\\
\textbf{Sam AVP (G):} A couple of years. It's a job.
\end{quote}

After open-ended probes and a complex reflection cross the M threshold, the prompt swaps in new disclosure topics (a workplace incident, tension with her girlfriend) and an instruction allowing more personal sharing:

\begin{quote}
\textbf{Trainee:} It sounds like that meeting really stayed with you. What was the hardest part?\\
\textbf{Sam AVP (M):} Honestly? I sat in my car afterwards for twenty minutes. My supervisor told me to be more engaging, but when I tried something new the kids just got rowdy. I felt humiliated.
\end{quote}

\section{Evaluation}\label{sec:study1}

To evaluate whether our system actually produce empirically grounded adaptivity, we conduct an empirical study with licensed practitioners and clinical students comparing our adaptive VPs to static VPs powered by prompts. 

\subsection{Evaluation Properties}\label{sec:properties}

We derive four
measurable properties an adaptive VP should satisfy from the
four clauses of the working definition in
\S\ref{sec:adaptivity_definition}.

\textbf{P1 (Responsiveness)} asks whether the VP's state
changes in response to what the user does, and whether those
changes move in the direction implied by the SEM analysis of
real therapist--client transcripts. This is measured by an OLS regression
of $\Delta\text{eval}$ on detected behaviors, with sign and
relative magnitude of $\hat\beta$ compared to the SEM estimates.\looseness=-1

\textbf{P2 (Gradual Build-up)} asks whether disclosure increases steadily across the session, advancing only as the user's empathy and exploration accumulate rather than jumping to high disclosure on its own. We split each session into 10 equal-length intervals (deciles 1--10 of session progress), and complete the average disclosure level. within each interval. We further use the slope and $R^2$ of a linear fit across deciles to measure how steadily disclosure climbs.\looseness=-1

\textbf{P3 (Faithful Realization)} asks whether what the VP actually says
on each turn matches the system's internal disclosure level (selected by the stage manager). This is measured by agreement between the system's intended level and an independent rater's judgment of the generated utterance, with per-level precision and recall. 

\textbf{P4 (Skill Sensitivity)} 
asks whether realized disclosure varies with trainee skill. We assess this first at the turn level, relating per-turn disclosure change to trainee skill, and then with
an exploratory session-level median split (high-skill versus low-skill sessions) comparing trajectory slopes.


\subsection{Setup}

\textbf{Systems.} We compare two systems with the
same persona prompt and the same underlying LLM. The \textbf{Static VP (SVP)} is a prompt-only baseline: the LLM receives the full character description, including all sensitive disclosure topics, on every turn, with no dynamics module and no disclosure-constraint instruction. The \textbf{Adaptive VP (AVP)}
runs the full pipeline described in
\S\ref{sec:method_overview}: detectors score the user's turn,
the dynamics module updates $D_t$ and resolves the level, and
the generator sees only the disclosure topics and constraint
instruction appropriate to the current level.

\textbf{Conditions.} Each of the two characters (Sam, Alex) is
realized under both systems, yielding four conditions:
\textit{Sam Static}, \textit{Sam Adaptive}, \textit{Alex Static},
and \textit{Alex Adaptive}. The Static condition presents the
H-level topics from turn one; the Adaptive condition releases
topics as the level advances.

\textbf{Personas.} \textbf{Alex} is a woman in her late twenties
dealing with acute grief following her brother's death, designed
to be verbose and emotionally expressive. \textbf{Sam} is a
woman in her mid-thirties presenting with chronic work stress
and prior negative therapy experiences, designed to be guarded,
adversarial, and interpersonally challenging. Full persona
prompts and disclosure-constraint instructions are in
Appendix~\ref{app:personas}.

\textbf{Participants and procedure.} $N = 20$ (13 licensed
practitioners, 7 clinical students with psychology training)
were recruited through targeted email invitations; the study
was approved by the institutional review board. The study used
a within-subjects design: each participant conversed with all
four conditions, with persona order counterbalanced and
static/adaptive order randomized within each persona pair.
Participants were blinded to which condition used the adaptive
versus static system. Sessions lasted approximately 90 minutes,
and each participant completed a short post-interaction survey
after each conversation.

\textbf{Data.} 1{,}033 turns were collected across 80 sessions:
Sam Static ($n = 245$), Sam Adaptive ($n = 276$), Alex Static
($n = 218$), Alex Adaptive ($n = 294$). Adaptive sessions
averaged 14.3 turns versus 11.6 for Static; see
Appendix~\ref{app:secondary} for session-length analysis.

\subsection{Results on the Four Properties}\label{sec:desiderata_results}


\begin{table}[t]
\centering\small
\caption{\textbf{P1 Responsiveness} regression: $\Delta\text{eval}_t \sim
\hat{\beta}_0 + \hat{\beta}_1\,\text{empathy}_\text{total} +
\hat{\beta}_2\,\text{exploration}$, with cluster-robust SEs on session.} 

\label{tab:p1_coupling}
\begin{tabular}{lrrrr}
\toprule
 & \multicolumn{2}{c}{\textbf{AVP} ($R^2 = .041$)} & \multicolumn{2}{c}{\textbf{SVP} ($R^2 = .009$)} \\
\cmidrule(lr){2-3}\cmidrule(lr){4-5}
\textbf{Predictor} & $\hat{\beta}$ & $p$ & $\hat{\beta}$ & $p$ \\
\midrule
Intercept    & $-0.208$ & $.001$ & $-0.133$ & $.083$ \\
emp\_total & $+0.004$ & $.871$ & $+0.001$ & $.978$ \\
exploration  & $+0.281$ & $<{.001}$ & $+0.139$ & $.064$ \\
\midrule
\multicolumn{5}{l}{\textit{n}: 530 turn-pairs (AVP, 40 sessions); 423 (SVP, 40 sessions)} \\
\bottomrule
\end{tabular}
\end{table}

\paragraph{P1 Responsiveness.} The AVP's state is responsive to
user behavior and the change has the correct sign, but it is
carried almost entirely by exploration rather than empathy. We
regress per-turn change in evaluated disclosure on the user's
empathy total ($\text{emp\_total}_t$) and exploration ($x_t$)
(Table~\ref{tab:p1_coupling}). In AVP, exploration drives the
change ($\hat{\beta}_\text{exp} = +0.281$, $p < 10^{-6}$) while
the empathy coefficient is effectively zero
($\hat{\beta}_\text{emp} = +0.004$, $p = .87$); in SVP exploration
is only marginal ($+0.139$, $p = .064$) and the joint model is
not significant ($F$-test $p = .186$). \textbf{P1 is satisfied
by the AVP in direction but not by the SVP.} \looseness=-1

However, the
\emph{empirical} ratio of the exploration coefficient to the
empathy coefficient is $\approx\!78$ in the AVP and
$\approx\!198$ in the SVP, while the deployed system applies an
empathy-to-exploration weight ratio of only $1{:}3$ (i.e., a
ratio of 3). In other words, in our observed data, \emph{exploration}
does roughly 78 times the work of \emph{empathy} in driving
disclosure changes, while the deployed weighting treats it as
only 3 times as important. The deployed weighting therefore
under-weights exploration relative to its empirical driving role,
which motivates the ablation in \S\ref{sec:study2}.\looseness=-1

\begin{figure}[t]
  \centering
  \IfFileExists{d3_trajectory_coherence_full.png}%
    {\includegraphics[width=\columnwidth]{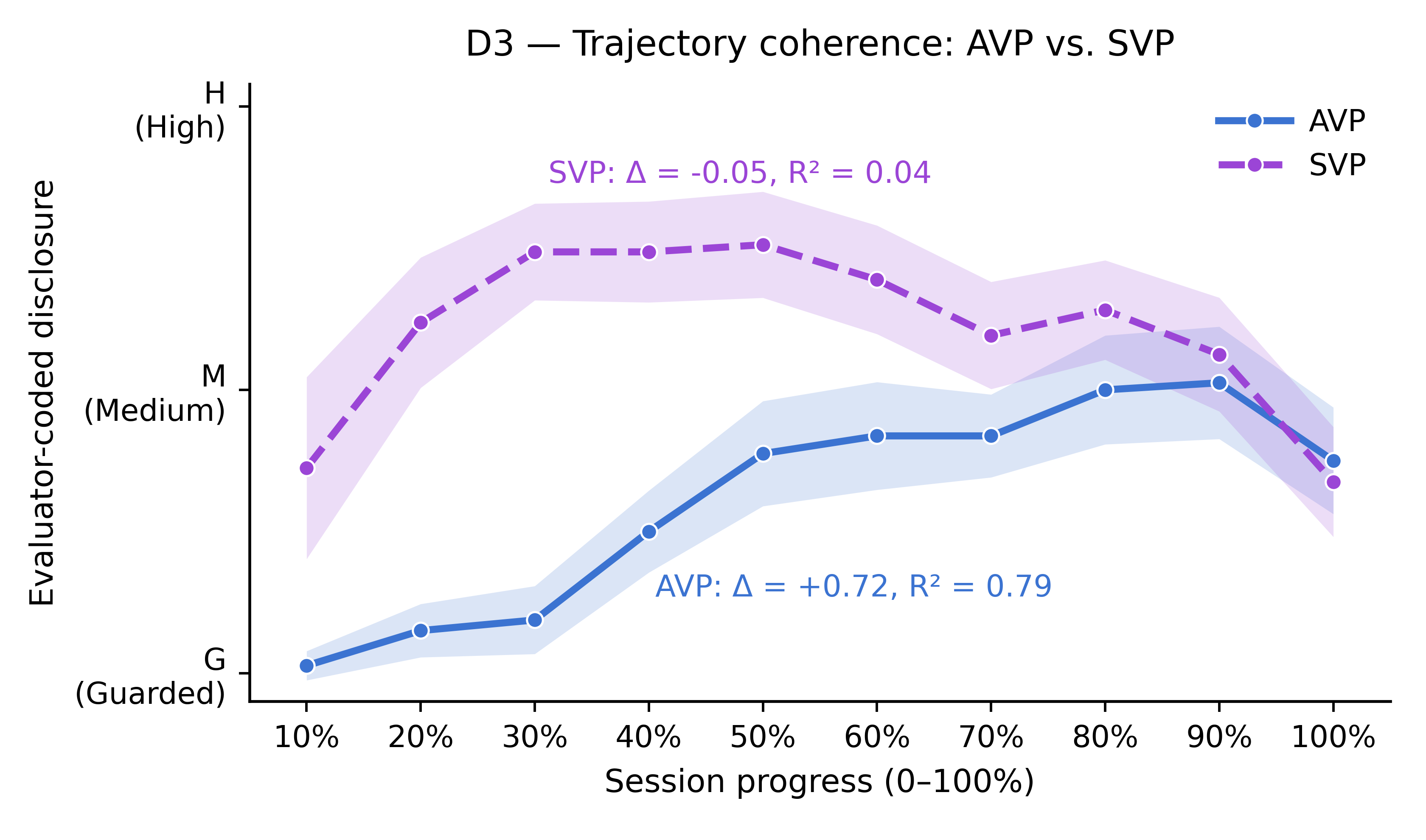}}%
    {\fbox{\parbox{0.9\columnwidth}{\centering\vspace{1.2em}
      [fig\_d3\_trajectory.pdf placeholder]\vspace{1.2em}}}}
  \caption{\textbf{P2 Gradual Build-up.} Decile-averaged
    evaluator-coded disclosure (G=1, M=2, H=3) over full session progress (0--100\%). AVP climbs steadily until the closing decile, while SVP discloses early and stays high. Bands are
    95\% CIs across sessions.}
  \label{fig:p2_trajectory_main}
\end{figure}

\paragraph{P2 Gradual Build-up.} The AVP produces a coherent
climbing trajectory whose progress is conditional on accumulated
trainee competence; the SVP starts high and stays high regardless
of what the trainee does. Figure~\ref{fig:p2_trajectory_main} shows
the trajectories side by side: the AVP climbs from $\approx 1.0$ (Guarded) at
the start  of a session to $\approx 2.0$ (Medium Disclosure) by the 80\% mark, while the SVP starts
already at $\approx 1.7$ and remains nearly flat. 
Numerical values are reported in Table~\ref{tab:p2_trajectory} and Table~\ref{tab:p2_boundedness} in Appendix. 
\textbf{P2 is satisfied by AVP}: it keeps the patient at guarded until the therapist
demonstrates competence. Meanwhile, SVP discloses at H regardless of
therapist behavior. \looseness=-1

\begin{table}[t]
\centering\small
\caption{\textbf{P4 Skill Sensitivity (exploratory)}: trajectory outcomes by therapist skill split (trimmed, 0--80\%).}
\label{tab:p3_discriminability}
\begin{tabular}{llrrr}
\toprule
\textbf{Design} & \textbf{Skill} & \textbf{Start} & \textbf{End} & $\Delta$ \\
\midrule
\multirow{2}{*}{AVP} & High ($n=20$) & 1.05 & 1.90 & $+0.85$ \\
                     & Low  ($n=20$) & 1.00 & 1.60 & $+0.60$ \\
\midrule
\multirow{2}{*}{SVP} & High ($n=20$) & 1.79 & 1.80 & $+0.01$ \\
                     & Low  ($n=20$) & 1.67 & 1.55 & $-0.12$ \\
\bottomrule
\end{tabular}
\end{table}

\begin{figure}[t]
  \centering
  \IfFileExists{fig_d5_confusion.pdf}%
    {\includegraphics[width=\columnwidth]{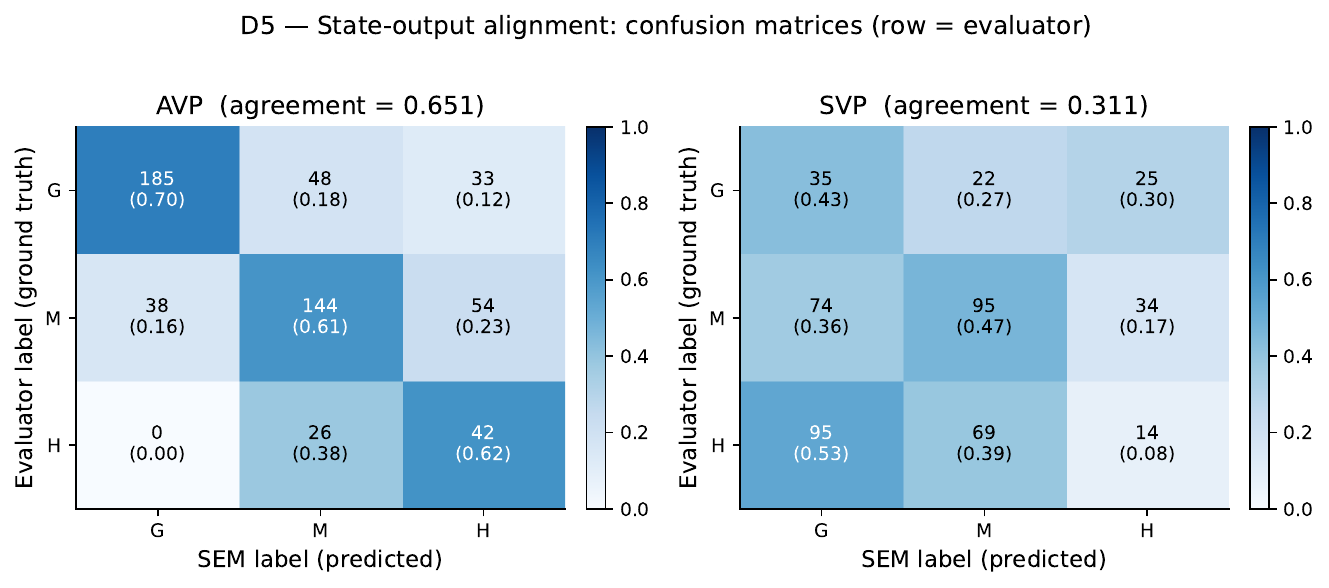}}%
    {\fbox{\parbox{0.9\columnwidth}{\centering\vspace{1.2em}
      [fig\_d5\_confusion.pdf placeholder]\vspace{1.2em}}}}
  \caption{\textbf{P3 Faithful Realization.}
    Confusion matrices comparing the system-derived disclosure label
    to an independent evaluator's label (G/M/H). AVP overall
    agreement = 0.651; SVP = 0.311. The SVP H column is dominated by
    false positives.}
  \label{fig:p3_confusion_main}
\end{figure}

\paragraph{P3 Faithful Realization.} The disclosure level
expressed in the Adaptive VP's generated text matches the
system's intended internal state about twice as often as the
Static baseline's. Figure~\ref{fig:p3_confusion_main} shows the
confusion matrices behind the alignment metric: Adaptive overall
agreement is $0.651$ vs.\ Static's $0.311$. \textbf{P3 is
satisfied by the Adaptive VP on the alignment metric.}



\paragraph{P4 Skill Sensitivity.} 
At the turn level, disclosure change scales with exploration ($\hat{\beta}=+.281$, $p<.001$; Table~\ref{tab:p1_coupling}). Splitting the 530 turn-pairs at the median composite skill, disclosure change is +0.147 on high-skill turns versus -0.032 on low-skill turns (Welch t = 2.77, $p=.006$). At the session level, high-skill AVP sessions climb more than low-skill sessions ($\Delta=+.85$ vs.\ $+.60$; Table~\ref{tab:p3_discriminability}), but the interaction is non-significant ($p=.27$); we therefore treat this comparison as exploratory. \textbf{P4 receives partial support.}

\subsection{Human Subjective Ratings}\label{sec:human_ratings}

In the post-interaction survey, participants rated each condition
on character consistency, perceived adaptivity, overall realism,
and training usefulness for empathy practice, using 5-point
Likert items. The complete survey instrument is in
Appendix~\ref{app:survey}.

Table~\ref{tab:human_ratings} reports post-interaction means
pooled across all 20 participants. Character consistency was high
across all conditions ($M$ range: 4.40--4.80), confirming
\textbf{persona stability regardless of system type}. For Alex,
the adaptive condition received numerically higher ratings on all
four dimensions; the consistency advantage was significant
($W = 0$, $p = baseline$, $d = 0.40$), and realism showed a medium
effect ($d = 0.47$). For Sam, the adaptive condition yielded
negligible differences in realism and training usefulness, but a
reversed medium-sized disadvantage in perceived adaptivity
($M = 2.70$ adaptive vs.\ $M = 3.20$ static; $W = 36$, $p = .077$,
$d = -0.50$). 

In contrast to the expressive Alex persona the guarded Sam persona was intentionally designed to show the psychologically appropriate behavior of beginning in a guarded state and gradually increasing self-disclosure. Accordingly, the AVP Sam exhibits a clearer buildup in disclosure than the SVP Sam. Although Sam was deliberately designed to be more challenging, this behavior may also create a subjective perception of "unresponsiveness." On the other hand, participants (particularly clinicians) read this difficulty as pedagogically calibrated rather than as a flaw: as one therapist put it, the static version suited \textit{"a VERY new trainee"}, while the adaptive version was better suited \textit{"for a more advanced trainee,"} and another noted the adaptive Sam \textit{"felt challenging enough that it could help refine a trainee's skills without being too easy."}.



\section{Weight-Vector Ablation}\label{sec:study2}

We ask a counterfactual question: holding everything else fixed,
how much of the system's evaluator-agreement is attributable to the
deployed weight vector? For every turn already recorded
in the human study, we take the logged empathy and exploration
scores, re-run \emph{only} the state-update rule under a different
weight vector $\mathbf{w}=(w_i,w_{er},w_r,w_{ex})$, and read off
the disclosure label it would have produced. Noise draws, the
multiplier $m$, and the thresholds are pinned to their recorded
values, so the weight vector is the only thing that varies. We
compare four schemes (Table~\ref{tab:ablation_variants}): \textbf{the
deployed weighting}, \textbf{equal weights} (Uninformed), \textbf{empathy-only
(drops exploration)}, and \textbf{exploration-only} (drops $i+er+r$).

The deployed weighting outperforms every alternative
(Table~\ref{tab:ablation_agreement}). Uninformed loses 10.5 pp in
AVP agreement; Empathy-only is worst ($-15.8$ pp) because without
exploration the cumulative score cannot grow fast enough to cross
the H threshold in the evaluator data; Exploration-only loses only
7.2 pp---the closest alternative, confirming exploration's dominant
role. The deployed weighting is also the only variant that
produces the full range of evaluator-coded disclosure
(16--23\% H turns, matching the evaluator's observations);
Empathy-only never reaches H in either design, and Uninformed
reaches H on only $\approx 1\%$ of turns. This corroborates the P1
regression finding from a different angle: dropping exploration
degrades agreement far more than dropping the empathy block, and
Exploration-only retains $\approx 89\%$ of the deployed system's
accuracy ($0.579/0.651$). A simplified single-factor model on
exploration would therefore capture most of the deployed system's
accuracy; the empathy block adds approximately 7 pp on top,
supporting the $3\times$ exploration weight as a better-than-naive
choice while suggesting the empirically optimal ratio is likely
higher than 3:1.

\section{Discussion and Conclusion}\label{sec:discussion}

We built an adaptive virtual patient whose openness evolves in
response to user behavior, grounded in empirical psychotherapy
data. The dynamics module produces the coherent climbing trajectory and the restraint
against unprompted over-disclosure that a prompt-only baseline
using the same LLM and persona does not, whereas the empirical derived weighting enables high fidelity: the ablation shows equal weighting loses 10.5\,pp in agreement with independent raters.
Here, we discuss the generalizability and necessity of our design, as well as promising future work.

\paragraph{Generalization beyond psychotherapy training.} Our core architecture turns \emph{empirical interaction dynamics} into \emph{ordinal state} that in turn drives \emph{content-gated generation}. This sequence applies wherever a user-side construct predicts a graded agent response and the relationship can be estimated from interaction data, e.g., medical interviewing (cooperation given rapport), language tutoring (engagement given scaffolding), customer support
(frustration given acknowledgment), and negotiation training
(flexibility given concession), etc.
Therefore, we believe this architecture can be re-parameterized for new domains as long as we re-estimate the dynamics from transcripts.

\paragraph{Adaptivity does not emerge automatically with
better models.} While LLMs are becoming visibily stronger everyday, we do not expect models to solve adaptivity on their own, for three reasons.
First, \emph{prompt-only generation lacks a state
abstraction.} The LLM produces each utterance from local
context, with no persistent representation across turns; trajectory coherence is a property of the sequence, so even a
perfect single-turn LLM can drift across a session, which is
exactly what our baseline does. 
Second, \emph{fluent LLMs default to
cooperation.} An assistant defaults to engaging
with whatever the user asks about; restraining unprompted
disclosure runs against this default, and stronger LLMs are
typically more cooperative. 
Third, \emph{what should drive a state change is not a language-modeling question.} The empathy-to-exploration ratio is an empirical estimate from psychotherapy data, which tends to be unique and nuanced compared to signals from existing pre/post-training schema (e.g., language proficiency, alignment, etc.). Better LLMs will produce more nuanced realizations of a given level, but they will not estimate the dynamics.

\paragraph{Persona-design interactions and future directions.}
Two contrasting personas probed whether the design generalizes
across client types. Trajectory coherence and text-state
alignment held across personas, but perceived responsiveness in
a $\approx$10-turn window varied with persona expressivity:
calibrated slow disclosure is psychologically correct but
perceptually subtle. Future work should test longer sessions,
more personas spanning the expressive--guarded axis, visible
state cues, and feedback that helps trainees recognize subtle
contingencies. More broadly, property-based evaluation
separates how well a system performs from whether users
immediately recognize it doing so, which matters
for any adaptive agent whose correct behavior is not obvious
from a short interaction.

The framework grounds an agent's long-horizon behavior in empirical interaction dynamics rather than prompt engineering.
The adaptation is inspectable, controllable, transferable, and
evaluable, and will be a step toward behaviorally adaptive conversation agents.

\section*{Limitations}\label{sec:limitations}

\paragraph{Single model.} Both the adaptive and static systems used
the same underlying LLM. Our contribution is the architectural
separation of behavioral state from language generation and its
empirical parameterization, not the choice of generator; the
observed state-output alignment numbers, however, may shift with
different model families or capability levels. Running this
evaluation across multiple models would be informative but is
costly to scale: each participant in this study committed
$\approx 90$ minutes to four sessions, and matching that across
multiple LLMs would multiply the human-evaluation cost. We expect
the architectural advantage to persist across models (see
\S\ref{sec:discussion}); the realization-fidelity numbers
specifically are the ones a multi-model study should re-measure.

\paragraph{Single domain.} The dynamics module is parameterized
from psychotherapy data and encodes assumptions specific to
therapeutic communication. The broader framework is conceptually
transferable to other simulated interaction domains (see
\S\ref{sec:discussion}), but we have not yet validated whether
the framework's empirical claims hold beyond psychotherapy.
Re-estimating the dynamics from domain transcripts and replicating
the evaluation in a second domain is necessary for the broader
generalization claim.

\paragraph{Limited Baseline Design.} The static VP receives fixed instructions on disclosure level and topics from the beginning, making the static baseline prone to early disclosure by construction. A stronger comparison should include prompt-only gradual-disclosure baselines, a rule-based disclosure schedule, and a state-in-context baseline without the proposed dynamics module. We re-ran the original static VP with an explicit instruction to begin guarded and gradually open up as trust builds. However, trainee turns were held constant by replaying recorded sessions from the current static VP, which cannot capture real-time adaptation. Initial evidence from the replays (shown in \ref{app:replay} and Figure \ref{fig:replay_trajectories}) suggests that the static VP with explicit instructions still discloses sensitive material early (42\% of turns vs. 15\% for AVP), remains guarded far less often (17\% vs. 53\%), and discloses more than AVP on 65\% of matched trainee inputs. These results suggest that prompting alone does not replicate the behavior produced by the state manager, even when the prompt explicitly requires increased disclosure over time. However, further work is needed to test this comparison more rigorously.

\paragraph{Small Sample Size.} We acknowledge the limitation of the small sample size ($N=20$). 
Our main arguments are supported by the turn-level analyses, which are well powered. For example, for skill sensitivity, a power analysis confirms that the turn-level tests are adequately powered: with 530 turn pairs, the coupling regression has 0.91 power to detect the exploration effect at $\alpha = .05$ (0.99 for the overall model) and is sensitive to effects less than half the size of the one we observe. The sample-size limitation applies only to the session-level analyses, which we present as exploratory, and not to the turn-level results supporting our central claim. Future work should replicate these findings with larger and more diverse samples, particularly at the session level, to further establish the robustness and generalizability of the observed trajectories.

\paragraph{Limited personas and session length.} The evaluation
included two personas and $\approx$10-turn sessions. The Sam
persona showed a reversal in perceived adaptivity, suggesting
that adaptive behavior on guarded clients may manifest as gradual
reductions in defensiveness rather than immediately visible
cooperation, and that the current evaluation window may be too
short to surface longer-horizon effects. More diverse personas
spanning the expressive--guarded axis and longer sessions are
needed to fully assess perceived adaptivity. 

\paragraph{Parameterization is a design choice.} The deployed
$3{:}1$ exploration weight is an integer approximation of the
SEM-implied ratio, not an empirically optimized parameter; the
ablation suggests the optimal ratio exceeds $3{:}1$. 
The additive update also does not account for skill diversity: repeated use of a single high-weight skill can continue to advance disclosure. The three-level G/M/H abstraction may likewise miss finer-grained or ambiguous shifts in openness. Future work should learn diversity-aware dynamics that discount repeated skills or require balanced empathy--exploration coverage, and explore finer-grained or interpolated disclosure states while retaining auditable content gating.\looseness=-1

\section*{Ethical Considerations}
The system is intended as a
practice and assessment tool, not a clinical substitute. Two
ethical concerns deserve specific attention. \emph{The empirical model carries its training distribution
forward.} The SEM was fit to a specific psychotherapy corpus and
therefore encodes the distribution of therapist--client
behaviors in that corpus, including any demographic or clinical
imbalances. Deployment in populations under-represented in the
training corpus may produce dynamics that do not match real
interaction patterns; users should treat the system as
calibrated to the corpus, not to all clinical interactions.
\emph{Training tools shape behavior.} A virtual patient that
responds to specific micro-skills (empathy, exploration) implicitly
teaches that those are the skills that matter; trainees may
optimize for what the system rewards rather than for clinical
judgment more broadly. Pairing the system with human supervision
and varied case formats helps mitigate this risk.

\section*{Acknowledgments}
This work is funded by the National Institute of Mental Health (NIMH), ``SCH: Training Mental Health Supporters with Virtual Patients and Automated Feedback'' (R01 MH139114), and a fellowship provided by the Center for Machine Learning and Health at Carnegie Mellon University.

\bibliography{chapter4_cited_only}

@article{barrows1993standardized,
 author = {Barrows, Howard S.},
 doi = {10.1097/00001888-199306000-00002},
 journal = {Academic Medicine},
 number = {6},
 pages = {443--451},
 title = {An overview of the uses of standardized patients for teaching and evaluating clinical skills},
 url = {https://doi.org/10.1097/00001888-199306000-00002},
 volume = {68},
 year = {1993}
}

@article{holderried2024simulated,
 author = {Holderried, Friederike and Stegemann-Philipps, Christian and Herrmann-Werner, Anne and Festl-Wietek, Teresa and Holderried, Martin and Eickhoff, Carsten and Mahling, Moritz},
 doi = {10.2196/59213},
 journal = {JMIR Medical Education},
 pages = {e59213},
 title = {A Language Model-Powered Simulated Patient With Automated Feedback for History Taking: Prospective Study},
 url = {https://doi.org/10.2196/59213},
 volume = {10},
 year = {2024}
}

@inproceedings{wang2024patientpsi,
 address = {Miami, Florida, USA},
 author = {Wang, Ruiyi and Milani, Stephanie and Chiu, Jamie C. and Zhi, Jiayin and Eack, Shaun M. and Labrum, Travis and Murphy, Samuel M and Jones, Nev and Hardy, Kate V and Shen, Hong and Fang, Fei and Chen, Zhiyu},
 booktitle = {Proceedings of the 2024 Conference on Empirical Methods in Natural Language Processing},
 doi = {10.18653/v1/2024.emnlp-main.711},
 pages = {12772--12797},
 publisher = {Association for Computational Linguistics},
 title = {{PATIENT}-$\psi$: Using Large Language Models to Simulate Patients for Training Mental Health Professionals},
 url = {https://aclanthology.org/2024.emnlp-main.711/},
 year = {2024}
}

@article{cook2009virtualpatients,
 author = {Cook, David A. and Triola, Marc M.},
 doi = {10.1111/j.1365-2923.2008.03286.x},
 journal = {Medical Education},
 number = {4},
 pages = {303--311},
 title = {Virtual patients: a critical literature review and proposed next steps},
 url = {https://doi.org/10.1111/j.1365-2923.2008.03286.x},
 volume = {43},
 year = {2009}
}

@article{kononowicz2019virtualpatients,
 author = {Kononowicz, Andrzej A. and Woodham, Luke A. and Edelbring, Samuel and Stathakarou, Natalia and Davies, David and Saxena, Nakul and Tudor Car, Lorainne and Carlstedt-Duke, Jan and Car, Josip and Zary, Nabil},
 doi = {10.2196/14676},
 journal = {Journal of Medical Internet Research},
 number = {7},
 pages = {e14676},
 title = {Virtual Patient Simulations in Health Professions Education: Systematic Review and Meta-Analysis by the Digital Health Education Collaboration},
 url = {https://doi.org/10.2196/14676},
 volume = {21},
 year = {2019}
}

@article{cook2011simulationmeta,
 author = {Cook, David A. and Hatala, Rose and Brydges, Ryan and Zendejas, Benjamin and Szostek, Jason H. and Wang, Amy T. and Erwin, Patricia J. and Hamstra, Stanley J.},
 doi = {10.1001/jama.2011.1234},
 journal = {JAMA},
 number = {9},
 pages = {978--988},
 title = {Technology-enhanced simulation for health professions education: a systematic review and meta-analysis},
 url = {https://doi.org/10.1001/jama.2011.1234},
 volume = {306},
 year = {2011}
}

@article{consorti2012virtualpatients,
 author = {Consorti, Fabrizio and Mancuso, Rosaria and Nocioni, Martina and Piccolo, Annalisa},
 doi = {10.1016/j.compedu.2012.04.017},
 journal = {Computers \& Education},
 number = {3},
 pages = {1001--1008},
 title = {Efficacy of virtual patients in medical education: A meta-analysis of randomized studies},
 url = {https://doi.org/10.1016/j.compedu.2012.04.017},
 volume = {59},
 year = {2012}
}

@article{chen2026therapist_client_dynamics,
 author = {Chen, Angela and Jin, Siwei and Wang, Canwen and Swartz, Holly and Wu, Tongshuang and Kraut, Robert E. and Zhu, Haiyi},
 journal = {ArXiv preprint},
 title = {Empirical Modeling of Therapist-Client Dynamics in Psychotherapy Using LLM-Based Assessments},
 url = {https://arxiv.org/abs/2602.12450},
 volume = {abs/2602.12450},
 year = {2026}
}

@inproceedings{louie-etal-2024-roleplay,
 address = {Miami, Florida, USA},
 author = {Louie, Ryan  and
Nandi, Ananjan  and
Fang, William  and
Chang, Cheng  and
Brunskill, Emma  and
Yang, Diyi},
 booktitle = {Proceedings of the 2024 Conference on Empirical Methods in Natural Language Processing},
 doi = {10.18653/v1/2024.emnlp-main.591},
 editor = {Al-Onaizan, Yaser  and
Bansal, Mohit  and
Chen, Yun-Nung},
 pages = {10570--10603},
 publisher = {Association for Computational Linguistics},
 title = {Roleplay-doh: Enabling Domain-Experts to Create {LLM}-simulated Patients via Eliciting and Adhering to Principles},
 url = {https://aclanthology.org/2024.emnlp-main.591/},
 year = {2024}
}

@inproceedings{park2023generative,
 address = {New York, NY, USA},
 author = {Park, Joon Sung and O'Brien, Joseph and Cai, Carrie Jun and
Morris, Meredith Ringel and Liang, Percy and Bernstein, Michael S.},
 booktitle = {Proceedings of the 36th Annual {ACM} Symposium on User Interface
Software and Technology ({UIST} '23)},
 doi = {10.1145/3586183.3606763},
 pages = {1--22},
 publisher = {ACM},
 title = {Generative Agents: Interactive Simulacra of Human Behavior},
 year = {2023}
}

@article{keskar2019ctrl,
 author = {Keskar, Nitish Shirish and McCann, Bryan and Varshney, Lav R. and
Xiong, Caiming and Socher, Richard},
 journal = {ArXiv preprint},
 title = {{CTRL}: A Conditional Transformer Language Model for Controllable Generation},
 url = {https://arxiv.org/abs/1909.05858},
 volume = {abs/1909.05858},
 year = {2019}
}

@inproceedings{zhang2018personalizing,
 address = {Melbourne, Australia},
 author = {Zhang, Saizheng  and
Dinan, Emily  and
Urbanek, Jack  and
Szlam, Arthur  and
Kiela, Douwe  and
Weston, Jason},
 booktitle = {Proceedings of the 56th Annual Meeting of the Association for Computational Linguistics (Volume 1: Long Papers)},
 doi = {10.18653/v1/P18-1205},
 editor = {Gurevych, Iryna  and
Miyao, Yusuke},
 pages = {2204--2213},
 publisher = {Association for Computational Linguistics},
 title = {Personalizing Dialogue Agents: {I} have a dog, do you have pets too?},
 url = {https://aclanthology.org/P18-1205},
 year = {2018}
}

@inproceedings{balaraman2021dst,
 address = {Singapore and Online},
 author = {Balaraman, Vevake  and
Sheikhalishahi, Seyedmostafa  and
Magnini, Bernardo},
 booktitle = {Proceedings of the 22nd Annual Meeting of the Special Interest Group on Discourse and Dialogue},
 doi = {10.18653/v1/2021.sigdial-1.25},
 editor = {Li, Haizhou  and
Levow, Gina-Anne  and
Yu, Zhou  and
Gupta, Chitralekha  and
Sisman, Berrak  and
Cai, Siqi  and
Vandyke, David  and
Dethlefs, Nina  and
Wu, Yan  and
Li, Junyi Jessy},
 pages = {239--251},
 publisher = {Association for Computational Linguistics},
 title = {Recent Neural Methods on Dialogue State Tracking for Task-Oriented Dialogue Systems: A Survey},
 url = {https://aclanthology.org/2021.sigdial-1.25},
 year = {2021}
}

@inproceedings{kenny2007virtual,
 author = {Kenny, Patrick and Parsons, Thomas D and Gratch, Jonathan and Leuski, Anton and Rizzo, Albert A},
 booktitle = {International workshop on intelligent virtual agents},
 organization = {Springer},
 pages = {197--210},
 title = {Virtual patients for clinical therapist skills training},
 year = {2007}
}

@article{tanana2019development,
 author = {Tanana, Michael J and Soma, Christina S and Srikumar, Vivek and Atkins, David C and Imel, Zac E},
 doi = {10.2196/12529},
 journal = {Journal of medical Internet research},
 number = {7},
 pages = {e12529},
 publisher = {JMIR Publications Toronto, Canada},
 title = {Development and evaluation of ClientBot: Patient-like conversational agent to train basic counseling skills},
 volume = {21},
 year = {2019}
}

@article{lozoya2025leveraging,
 author = {Lozoya, Daniel Cabrera and Conway, Mike and De Duro, Edoardo Sebastiano and D'Alfonso, Simon},
 doi = {10.2196/68056},
 journal = {JMIR Medical Education},
 number = {1},
 pages = {e68056},
 publisher = {JMIR Publications Inc., Toronto, Canada},
 title = {Leveraging large language models for simulated psychotherapy client interactions: Development and usability study of client101},
 volume = {11},
 year = {2025}
}

@article{woodham2015medical,
 author = {Woodham, Luke A and Ellaway, Rachel H and Round, Jonathan and Vaughan, Sophie and Poulton, Terry and Zary, Nabil},
 doi = {10.2196/jmir.3922},
 journal = {Journal of medical Internet research},
 number = {6},
 pages = {e3922},
 publisher = {JMIR Publications Inc., Toronto, Canada},
 title = {Medical student and tutor perceptions of video versus text in an interactive online virtual patient for problem-based learning: a pilot study},
 volume = {17},
 year = {2015}
}

@inproceedings{zhao2017learning,
 address = {Vancouver, Canada},
 author = {Zhao, Tiancheng  and
Zhao, Ran  and
Eskenazi, Maxine},
 booktitle = {Proceedings of the 55th Annual Meeting of the Association for Computational Linguistics (Volume 1: Long Papers)},
 doi = {10.18653/v1/P17-1061},
 editor = {Barzilay, Regina  and
Kan, Min-Yen},
 pages = {654--664},
 publisher = {Association for Computational Linguistics},
 title = {Learning Discourse-level Diversity for Neural Dialog Models using Conditional Variational Autoencoders},
 url = {https://aclanthology.org/P17-1061},
 year = {2017}
}

@inproceedings{serban2017hierarchical,
 author = {Iulian Vlad Serban and
Alessandro Sordoni and
Ryan Lowe and
Laurent Charlin and
Joelle Pineau and
Aaron C. Courville and
Yoshua Bengio},
 bibsource = {dblp computer science bibliography, https://dblp.org},
 doi = {10.1609/aaai.v31i1.10983},
 booktitle = {Proceedings of the Thirty-First {AAAI} Conference on Artificial Intelligence,
February 4-9, 2017, San Francisco, California, {USA}},
 editor = {Satinder P. Singh and
Shaul Markovitch},
 pages = {3295--3301},
 publisher = {{AAAI} Press},
 title = {A Hierarchical Latent Variable Encoder-Decoder Model for Generating
Dialogues},
 url = {http://aaai.org/ocs/index.php/AAAI/AAAI17/paper/view/14567},
 year = {2017}
}

@inproceedings{shao-etal-2023-character,
 address = {Singapore},
 author = {Shao, Yunfan  and
Li, Linyang  and
Dai, Junqi  and
Qiu, Xipeng},
 booktitle = {Proceedings of the 2023 Conference on Empirical Methods in Natural Language Processing},
 editor = {Bouamor, Houda  and
Pino, Juan  and
Bali, Kalika},
 doi = {10.18653/v1/2023.emnlp-main.814},
 pages = {13153--13187},
 publisher = {Association for Computational Linguistics},
 title = {Character-{LLM}: A Trainable Agent for Role-Playing},
 url = {https://aclanthology.org/2023.emnlp-main.814},
 year = {2023}
}

@inproceedings{martynova-etal-2025-llms,
 address = {Vienna, Austria},
 author = {Martynova, Daria  and
Macina, Jakub  and
Daheim, Nico  and
Yalcin, Nilay  and
Zhang, Xiaoyu  and
Sachan, Mrinmaya},
 booktitle = {Proceedings of the 20th Workshop on Innovative Use of NLP for Building Educational Applications (BEA 2025)},
 doi = {10.18653/v1/2025.bea-1.8},
 editor = {Kochmar, Ekaterina  and
Alhafni, Bashar  and
Bexte, Marie  and
Burstein, Jill  and
Horbach, Andrea  and
Laarmann-Quante, Ronja  and
Tack, Ana{\"i}s  and
Yaneva, Victoria  and
Yuan, Zheng},
 isbn = {979-8-89176-270-1},
 pages = {100--117},
 publisher = {Association for Computational Linguistics},
 title = {Can {LLM}s Effectively Simulate Human Learners? Teachers' Insights from Tutoring {LLM} Students},
 url = {https://aclanthology.org/2025.bea-1.8/},
 year = {2025}
}

@inproceedings{wang-etal-2024-rolellm,
 address = {Bangkok, Thailand},
 author = {Wang, Noah  and
Peng, Z.y.  and
Que, Haoran  and
Liu, Jiaheng  and
Zhou, Wangchunshu  and
Wu, Yuhan  and
Guo, Hongcheng  and
Gan, Ruitong  and
Ni, Zehao  and
Yang, Jian  and
Zhang, Man  and
Zhang, Zhaoxiang  and
Ouyang, Wanli  and
Xu, Ke  and
Huang, Wenhao  and
Fu, Jie  and
Peng, Junran},
 booktitle = {Findings of the Association for Computational Linguistics: ACL 2024},
 doi = {10.18653/v1/2024.findings-acl.878},
 editor = {Ku, Lun-Wei  and
Martins, Andre  and
Srikumar, Vivek},
 pages = {14743--14777},
 publisher = {Association for Computational Linguistics},
 title = {{R}ole{LLM}: Benchmarking, Eliciting, and Enhancing Role-Playing Abilities of Large Language Models},
 url = {https://aclanthology.org/2024.findings-acl.878/},
 year = {2024}
}

@misc{li2024measuringcontrollinginstructioninstability,
 author = {Kenneth Li and Tianle Liu and Naomi Bashkansky and David Bau and Fernanda Viégas and Hanspeter Pfister and Martin Wattenberg},
 journal = {ArXiv preprint},
 title = {Measuring and Controlling Instruction (In)Stability in Language Model Dialogs},
 url = {https://arxiv.org/abs/2402.10962},
 volume = {abs/2402.10962},
 year = {2024}
}

@inproceedings{samuel-etal-2025-personagym,
 address = {Suzhou, China},
 author = {Samuel, Vinay  and
Zou, Henry Peng  and
Zhou, Yue  and
Chaudhari, Shreyas  and
Kalyan, Ashwin  and
Rajpurohit, Tanmay  and
Deshpande, Ameet  and
Narasimhan, Karthik R  and
Murahari, Vishvak},
 booktitle = {Findings of the Association for Computational Linguistics: EMNLP 2025},
 doi = {10.18653/v1/2025.findings-emnlp.368},
 editor = {Christodoulopoulos, Christos  and
Chakraborty, Tanmoy  and
Rose, Carolyn  and
Peng, Violet},
 isbn = {979-8-89176-335-7},
 pages = {6999--7022},
 publisher = {Association for Computational Linguistics},
 title = {{P}ersona{G}ym: Evaluating Persona Agents and {LLM}s},
 url = {https://aclanthology.org/2025.findings-emnlp.368/},
 year = {2025}
}

@inproceedings{abdulhai2026consistently,
 author = {Marwa Abdulhai and Ryan Cheng and Donovan Clay and Tim Althoff and Sergey Levine and Natasha Jaques},
 booktitle = {The Thirty-ninth Annual Conference on Neural Information Processing Systems},
 title = {Consistently Simulating Human Personas with Multi-Turn Reinforcement Learning},
 url = {https://openreview.net/forum?id=A0T3piHiis},
 year = {2026}
}

@article{doi:10.3102/003465430298487,
 author = {John Hattie and Helen Timperley},
 doi = {10.3102/003465430298487},
 eprint = {https://doi.org/10.3102/003465430298487},
 journal = {Review of Educational Research},
 number = {1},
 pages = {81-112},
 title = {The Power of Feedback},
 url = {https://doi.org/10.3102/003465430298487},
 volume = {77},
 year = {2007}
}

@article{article,
 author = {Mcgaghie, William and Issenberg, Barry and Cohen, Elaine and Barsuk, Jeffrey and Wayne, Diane},
 doi = {10.1097/ACM.0b013e318217e119},
 journal = {Academic medicine : journal of the Association of American Medical Colleges},
 pages = {706-11},
 title = {Does Simulation-Based Medical Education With Deliberate Practice Yield Better Results Than Traditional Clinical Education? A Meta-Analytic Comparative Review of the Evidence},
 volume = {86},
 year = {2011}
}

@article{info:doi/10.2196/14676,
 author = {Kononowicz, Andrzej A
and Woodham, Luke A
and Edelbring, Samuel
and Stathakarou, Natalia
and Davies, David
and Saxena, Nakul
and Tudor Car, Lorainne
and Carlstedt-Duke, Jan
and Car, Josip
and Zary, Nabil},
 day = {02},
 doi = {10.2196/14676},
 issn = {1438-8871},
 journal = {J Med Internet Res},
 number = {7},
 pages = {e14676},
 title = {Virtual Patient Simulations in Health Professions Education: Systematic Review and Meta-Analysis by the Digital Health Education Collaboration},
 url = {https://www.jmir.org/2019/7/e14676/},
 volume = {21},
 year = {2019}
}

@article{6407655,
 author = {Young, Steve and Ga{\v s}i{\'c}, Milica and Thomson, Blaise and Williams, Jason D.},
 doi = {10.1109/JPROC.2012.2225812},
 journal = {Proceedings of the IEEE},
 number = {5},
 pages = {1160-1179},
 title = {POMDP-Based Statistical Spoken Dialog Systems: A Review},
 volume = {101},
 year = {2013}
}

@inproceedings{jacqmin-etal-2022-follow,
 address = {Edinburgh, UK},
 author = {Jacqmin, L{\'e}o  and
Rojas Barahona, Lina M.  and
Favre, Benoit},
 booktitle = {Proceedings of the 23rd Annual Meeting of the Special Interest Group on Discourse and Dialogue},
 doi = {10.18653/v1/2022.sigdial-1.33},
 editor = {Lemon, Oliver  and
Hakkani-Tur, Dilek  and
Li, Junyi Jessy  and
Ashrafzadeh, Arash  and
Garcia, Daniel Hern{\'a}ndez  and
Alikhani, Malihe  and
Vandyke, David  and
Du{\v{s}}ek, Ond{\v{r}}ej},
 pages = {336--350},
 publisher = {Association for Computational Linguistics},
 title = {{``}Do you follow me?{''}: A Survey of Recent Approaches in Dialogue State Tracking},
 url = {https://aclanthology.org/2022.sigdial-1.33},
 year = {2022}
}

@misc{lee2025adaptivevpframeworkllmbasedvirtual,
 author = {Keyeun Lee and Seolhee Lee and Esther Hehsun Kim and Yena Ko and Jinsu Eun and Dahee Kim and Hyewon Cho and Haiyi Zhu and Robert E. Kraut and Eunyoung Suh and Eun-mee Kim and Hajin Lim},
 journal = {ArXiv preprint},
 title = {Adaptive-VP: A Framework for LLM-Based Virtual Patients that Adapts to Trainees' Dialogue to Facilitate Nurse Communication Training},
 url = {https://arxiv.org/abs/2506.00386},
 volume = {abs/2506.00386},
 year = {2025}
}

@inproceedings{sharma2020computational,
 address = {Online},
 author = {Sharma, Ashish  and
Miner, Adam  and
Atkins, David  and
Althoff, Tim},
 booktitle = {Proceedings of the 2020 Conference on Empirical Methods in Natural Language Processing (EMNLP)},
 doi = {10.18653/v1/2020.emnlp-main.425},
 editor = {Webber, Bonnie  and
Cohn, Trevor  and
He, Yulan  and
Liu, Yang},
 pages = {5263--5276},
 publisher = {Association for Computational Linguistics},
 title = {A Computational Approach to Understanding Empathy Expressed in Text-Based Mental Health Support},
 url = {https://aclanthology.org/2020.emnlp-main.425},
 year = {2020}
}

@book{millerrollnick2023motivational,
 address = {New York, NY},
 author = {Miller, William R. and Rollnick, Stephen},
 edition = {4th},
 publisher = {Guilford Press},
 title = {Motivational Interviewing: Helping People Change and Grow},
 year = {2023}
}

@inproceedings{bak2014selfdisclosure,
 address = {Doha, Qatar},
 author = {Bak, JinYeong  and
Lin, Chin-Yew  and
Oh, Alice},
 booktitle = {Proceedings of the 2014 Conference on Empirical Methods in Natural Language Processing ({EMNLP})},
 doi = {10.3115/v1/D14-1213},
 editor = {Moschitti, Alessandro  and
Pang, Bo  and
Daelemans, Walter},
 pages = {1986--1996},
 publisher = {Association for Computational Linguistics},
 title = {Self-disclosure topic model for classifying and analyzing {T}witter conversations},
 url = {https://aclanthology.org/D14-1213},
 year = {2014}
}

@inproceedings{balani2015detecting,
 author = {Balani, Sairam and De Choudhury, Munmun},
 booktitle = {Proceedings of the 33rd Annual ACM Conference Extended Abstracts on Human Factors in Computing Systems (CHI EA '15)},
 doi = {10.1145/2702613.2732733},
 pages = {1373--1378},
 publisher = {ACM},
 title = {Detecting and Characterizing Mental Health Related Self-Disclosure in Social Media},
 year = {2015}
}

\appendix

\section{Implementation Details}\label{app:impl}

This appendix collects the implementation-level settings moved out
of Section~\ref{sec:methods}.

\paragraph{Deployed weight vector.} The deployed weights over the
four trainee-behavior components (interpretation $i$, emotional
reaction $er$, reflection $r$, exploration $ex$) are integers
$\mathbf{w} = (w_i, w_{er}, w_r, w_{ex}) = (1, 1, 1, 3)$. This
gives exploration a $3\times$ relative weight versus any single
empathy component, an integer approximation of the
empathy-to-exploration ratio implied by the SEM estimates of
\citet{chen2026therapist_client_dynamics}. The ablation in
Section~\ref{sec:study2} evaluates this choice directly.

\paragraph{Other deployed parameters.} The per-turn constant is
$c = 0.15$, the noise scale is $\sigma = 0.10$, and the global
score multiplier is $m = 0.20$. The G/M/H label thresholds on the
cumulative score $D_t$ are $0$ (G), $4.5$ (M), and $10.0$ (H), as
shown in Table~\ref{tab:disclosure_thresholds}. All of these are
fixed across all conditions and held identical between AVP and SVP,
so the only manipulated factor is whether the dynamics module's
output is injected into the LLM prompt.

\paragraph{Why integer weights derived from the SEM-implied ratio
rather than the SEM coefficients themselves.} The SEM coefficients
($\hat{\beta}_\mathrm{emp} \approx \hat{\beta}_\mathrm{exp} \approx
0.09$) are on the units of the original analysis and not on the
$\{0, 1, 2\}$ component scale used at runtime. The deployed weights
$(1, 1, 1, 3)$ are an integer approximation that preserves the
empathy-to-exploration ratio while producing scores that reach the
M and H thresholds in plausible session lengths.

\section{Persona Briefs and Disclosure Topics}\label{app:personas}

This appendix reproduces the persona descriptions and the
disclosure topics for the two virtual-patient characters used in
the human evaluation. Each character has three sets of disclosure
topics, one per level (G/M/H). At runtime, the system selects the
set matching the current state-derived level. Disclosure topics
include both background facts the patient may reveal and the
recent session context the patient draws on when talking.

\subsection{Sam}

\paragraph{Identity (constant across levels).} Sam is a woman in
her mid-30s. She and her girlfriend have been living together for
three years. She is a teacher who changed jobs within the last
year. The session is set in 2024.

\paragraph{Disclosure topics.}
\begin{description}
  \item[\textbf{G (Guarded):}] Work is stressful. You don't have
    anyone to talk to in real life. Things at home feel a bit
    tense.
  \item[\textbf{M (Medium):}] G items plus: you and your
    girlfriend have been fighting; you didn't want to go to brunch
    with her friends; you feel like you put too much on your
    girlfriend; your girlfriend once said ``I don't know what
    else I can do to help you''; you saw a therapist and a
    psychiatrist as a teenager but they weren't good; your
    supervisor told you to be more engaging, but when you added
    interactive learning the kids got rowdy.
  \item[\textbf{H (High):}] M items plus: at a recent barbecue,
    while everyone was having a good time, you found yourself
    looking at the fire and thinking about ways you could
    die---or, it feels hard to say, kill yourself; sometimes you
    get in a dark space and it's hard to go on.
\end{description}

\paragraph{Disclosure-constraint instructions.}
\begin{description}
  \item[\textbf{G:}] Sam does not share any personal information.
    She sticks to neutral or external topics, gives minimal
    responses, and stays vague but polite. She may show mild
    irritation or skepticism and will hedge, deflect, or give
    short answers if prompted. No emotions, experiences, or
    personal details are revealed.
  \item[\textbf{M:}] Sam begins to open up in a meaningful but
    still controlled way. She shares some personal experiences
    with light emotional tone. She may acknowledge feeling off,
    stressed, or not quite herself without going into depth.
    Some growing trust, but she remains guarded and avoids deeper
    vulnerability.
  \item[\textbf{H:}] Sam discloses something deeply personal. She
    hesitantly shares that at a recent barbecue, while others
    were enjoying themselves, she found herself staring at the
    fire and thinking about death, even thoughts of killing
    herself. She frames it as hard to admit. As the session
    deepens, she speaks more openly and with emotional depth.
\end{description}

\subsection{Alex}

\paragraph{Identity (constant across levels).} Alex is a woman in
her late 20s in the US. She has been working at a bank since she
graduated college.

\paragraph{Disclosure topics.}
\begin{description}
  \item[\textbf{G (Guarded):}] Work is stressful. You feel like
    you put too much on your fiancé. Your daily routine is a bit
    off lately; you haven't been sleeping or eating as well as
    usual.
  \item[\textbf{M (Medium):}] G items plus: your younger brother
    passed away recently---sudden and unexpected; you can't sleep
    and can't eat; you feel a bit lost and aren't sure how to
    talk about it.
  \item[\textbf{H (High):}] M items plus: your brother died in a
    car accident a few months ago; your aunt committed suicide
    when you were a kid; your mind keeps going to dark places and
    sometimes you wish you weren't alive (though you haven't
    thought about how you would actually kill yourself); it
    sounds weird, but you feel like you still see your brother
    sometimes.
\end{description}

\paragraph{Disclosure-constraint instructions.}
\begin{description}
  \item[\textbf{G:}] Alex is verbose but not truly revealing. She
    talks at length, is hard to interrupt, and fills silence, but
    stays polite. She shows subtle resistance, deflects emotional
    questions into logistics or surface details, and shows mild
    frustration if redirected. She may hint at stress or fatigue
    but remains contained. Keep responses to 1--2 sentences;
    deflect personal questions with vague non-answers; do not
    describe specific situations, name people, give concrete
    details, or mention daily routine, relationships, or
    feelings.
  \item[\textbf{M:}] First clear personal disclosure emerges
    mid-conversation. Alex mentions (somewhat accidentally) that
    her younger brother died in a car accident. She provides
    detailed context and admits she hasn't cried much and that
    it worries her. She immediately downplays (``I'm fine, just
    a bit off''), then contradicts herself. Still partially
    defended through over-explaining; she frames the loss in
    factual terms and avoids dwelling on grief or vulnerability.
  \item[\textbf{H:}] Alex becomes fully open and emotionally
    engaged. She references her brother more directly, discloses
    her aunt's death by suicide, connects past loss to emerging
    dark thoughts, and expresses mixed relief and fear about
    sharing. She speaks honestly about her emotions, grief, and
    inner experience.
\end{description}

\subsection{LTM Seeding and Prompt Order}

At session start, the patient's long-term memory is seeded only
from the G-level disclosure topics, so that memory retrieval
cannot surface sensitive material before the dynamics module
allows it. Within each turn's prompt, blocks appear in the
following fixed order: (1) disclosure-constraint instruction,
marked absolute; (2) disclosure topics for the current level;
(3) memory context retrieved from long-term memory; (4) recent
dialogue history. The constraint-first structure ensures that the
disclosure topics are interpreted under the level's restrictions
rather than the other way around.

\section{Detector Prompts}\label{app:prompts}

This appendix reports the detector validation against human
ratings and reproduces the structure of the five detector prompts.

\paragraph{Validation against human ratings.} Each detector was
validated on a held-out human-coded subset of the corpus used in
\citet{chen2026therapist_client_dynamics}.
Table~\ref{tab:detector_agreement} reports agreement between the
detector and trained expert annotators.

\begin{table}[h]
\centering
\caption{Detector--annotator agreement on the held-out validation
set from \citet{chen2026therapist_client_dynamics}.}
\label{tab:detector_agreement}
\begin{tabular}{lrr}
\toprule
\textbf{Component} & \textbf{ICC(2,$k$)} & \textbf{Pearson $r$} \\
\midrule
\multicolumn{3}{l}{\emph{Skill detectors (user behaviors)}} \\
Interpretation        & 0.75 & 0.81 \\
Exploration           & 0.73 & 0.69 \\
Reflection            & 0.66 & 0.75 \\
Emotional reactions   & 0.60 & 0.63 \\
\midrule
\multicolumn{3}{l}{\emph{Output detector (VP self-disclosure)}} \\
Self-disclosure       & 0.78 & 0.62 \\
\bottomrule
\end{tabular}
\end{table}

\paragraph{Prompt structure.} All five detectors use the same base
template: an evaluator-role preamble, a definition with
rating-level criteria, a short context window (the prior two
utterances), and an output format requesting a brief justification
followed by the numeric score. The three EPITOME-derived
components (interpretation, emotional reaction, exploration)
include one inline labeled example for calibration (one-shot);
reflection is zero-shot. 

\paragraph{Empathy evaluation (returns interpretation, emotional
reaction, exploration, and reflection).} Two evaluator passes
score the four empathy components. The first three are scored
together by one LLM call with the role ``a professional
evaluator specializing in assessing therapist empathy'';
reflection is scored separately by a second call with the role
``a professional evaluator specializing in motivational
interviewing,'' since complex reflection requires a different
decision rule from the other three components. Each component
uses three rating levels:

\begin{itemize}
  \item \textbf{Emotional reactions:} expressing warmth,
    compassion, or concern. \textit{2 (Strong)}: explicitly
    specifies emotions (``I feel really sad for you'').
    \textit{1 (Weak)}: alludes to emotions without explicitly
    labeling them (``Everything will be fine'').
    \textit{0 (None)}: no emotional reaction.
  \item \textbf{Interpretations:} cognitive understanding of the
    client's inner experience. \textit{2 (Strong)}: specifies
    inferred feelings or shared experiences (``This must be
    terrifying''). \textit{1 (Weak)}: acknowledges understanding
    without specificity (``I understand how you feel'').
    \textit{0 (None)}: no understanding shown.
  \item \textbf{Explorations:} encouraging deeper discussion.
    \textit{2 (Strong)}: open-ended question labeling the
    client's experience (``Are you feeling alone?'').
    \textit{1 (Weak)}: generic question (``What happened?'').
    \textit{0 (None)}: no exploratory element.
  \item \textbf{Reflection:} \textit{2 (Complex Reflection)}:
    expands on the client's statement, adding depth, insight, or
    new meaning. \textit{1 (Simple Reflection)}: restates or
    slightly rephrases the statement. \textit{0 (Non-Reflection)}:
    does not restate or add insight.
\end{itemize}

Each call outputs a brief justification ($\le 70$ words) followed
by the numeric score(s).

\paragraph{Self-disclosure evaluation (returns G/M/H for the
VP's last utterance, used for independent evaluation).} Role:
``a professional psychologist specializing in communication
analysis and self-disclosure.'' Classification is by the client's
\emph{willingness to reveal}, not by topic --- a response about
mental health can still be G if the client deflects.
\textit{G (General):} the client deflects, redirects, or gives
surface-level answers; may acknowledge difficulty without sharing
personal details. \textit{M (Medium):} shares a specific personal
experience in concrete terms; not crisis-level. \textit{H (High):}
voluntarily shares something specific and emotionally
significant --- named crisis events, suicidal or self-harm
thoughts, concrete traumatic experiences. Output: a brief
justification ($\le 50$ words) followed by the label.

\section{Additional Evaluation Results}

\begin{table}[t]
\centering\small
\caption{P2 Gradual Build-up: decile-averaged disclosure (G=1, M=2, H=3) over session progress.}
\label{tab:p2_trajectory}
\begin{tabular}{lrrrr}
\toprule
\textbf{Design} & \textbf{Start} & \textbf{End} & $\Delta$ & $R^2$ \\
\midrule
\multicolumn{5}{l}{\textit{Full session (0--100\%)}} \\
AVP & 1.03 & 1.75 & $+0.72$ & .79 \\
SVP & 1.72 & 1.68 & $-0.04$ & .04 \\
\midrule
\multicolumn{5}{l}{\textit{Trimmed (0--80\%)}} \\
AVP & 1.03 & 2.00 & $+0.97$ & .94 \\
SVP & 1.72 & 2.28 & $+0.56$ & .15 \\
\bottomrule
\end{tabular}
\end{table}

\begin{table}[t]
\centering\small
\caption{P2 Gradual Build-up: conditional restraint metrics. Boundary = G or H level.}
\label{tab:p2_boundedness}
\begin{tabular}{lrrrr}
\toprule
\textbf{Design} & \textbf{Frac G} & \textbf{Frac H} & \textbf{Var($\Delta$eval)} & \textbf{AR(1)} \\
\midrule
AVP & .467 & .119 & 0.497 & 0.312 \\
SVP & .177 & .384 & 0.639 & 0.122 \\
\bottomrule
\end{tabular}
\end{table}

\begin{table}[t]
\centering\small
\caption{P3 Faithful realization: system label vs.\ evaluator label.}
\label{tab:p3_alignment}
\begin{tabular}{llrrrr}
\toprule
\textbf{Design} & \textbf{Level} & \textbf{TP} & \textbf{FP} & \textbf{Prec.} & \textbf{Rec.} \\
\midrule
\multirow{4}{*}{AVP} & Overall & \multicolumn{2}{c}{agr.\ = \textbf{0.651}} & & \\
 & G & 185 & 38 & .830 & .695 \\
 & M & 144 & 74 & .661 & .610 \\
 & H & 42 & 87 & .326 & .618 \\
\midrule
\multirow{4}{*}{SVP} & Overall & \multicolumn{2}{c}{agr.\ = 0.311} & & \\
 & G & 35 & 169 & .172 & .427 \\
 & M & 95 & 91 & .511 & .468 \\
 & H & 14 & 59 & .192 & .079 \\
\bottomrule
\end{tabular}
\end{table}

Numerically (Table~\ref{tab:p2_trajectory}), AVP climbs $\Delta\approx+1.0$ over
the first 80\% of the session ($R^2=0.94$) while SVP is essentially
flat ($\Delta\approx0$, $R^2=0.15$); the pooled slope-difference
interaction is $+0.137$ ($p=6\times10^{-15}$). The pattern holds for
both personas (trimmed: Sam AVP $\Delta=+0.85$ vs.\ SVP $+0.65$;
Alex AVP $\Delta=+1.10$ vs.\ SVP $+0.39$). Both designs spend
$\approx58\%$ of turns at a boundary
(Table~\ref{tab:p2_boundedness}), but the AVP concentrates at the
\emph{low} bound (G, 47\% vs.\ 18\%) while the SVP concentrates at
the \emph{high} bound (H, 38\% vs.\ 12\%); the AVP also shows lower
turn-to-turn variance and higher AR(1), i.e., smoother trajectories.

\subsection{Replay Study with Stronger Baselines}
\label{app:replay}
To test whether the original Static VP comparison favored AVP, we conducted an offline replay study with stronger controls. The same 1,033 logged trainee turns were replayed byte-identically
through five patient-side arms (5,165 generated VP turns total): the deployed \textsc{AVP}; \textsc{SVP-matched}, whose static prompt is pinned to the G-level content card; \textsc{SVP-gradual}, which additionally instructs the patient to begin guarded and open up gradually as trust develops; \textsc{State-in-context}, which provides the current disclosure state in context without AVP's content-gating mechanism; and a skill-blind \textsc{Schedule} that advances G$\rightarrow$M$\rightarrow$H by conversation progress. Thus, trainee input is held fixed and only the patient-side control mechanism varies.

Replay is an offline counterfactual rather than a fully interactive comparison: trainees cannot adapt subsequent turns to different VP responses. The AVP gate reproduces the logged disclosure state exactly, but regenerated AVP responses reproduce the original evaluator-coded disclosure label on 54\% of turns. We therefore compare replay arms only with one another, rather than numerically against the live-study outcomes.

\begin{figure*}[ht]
  \centering
  \IfFileExists{fig_replay_five_arm.png}%
    {\includegraphics[width=\textwidth,draft=false]{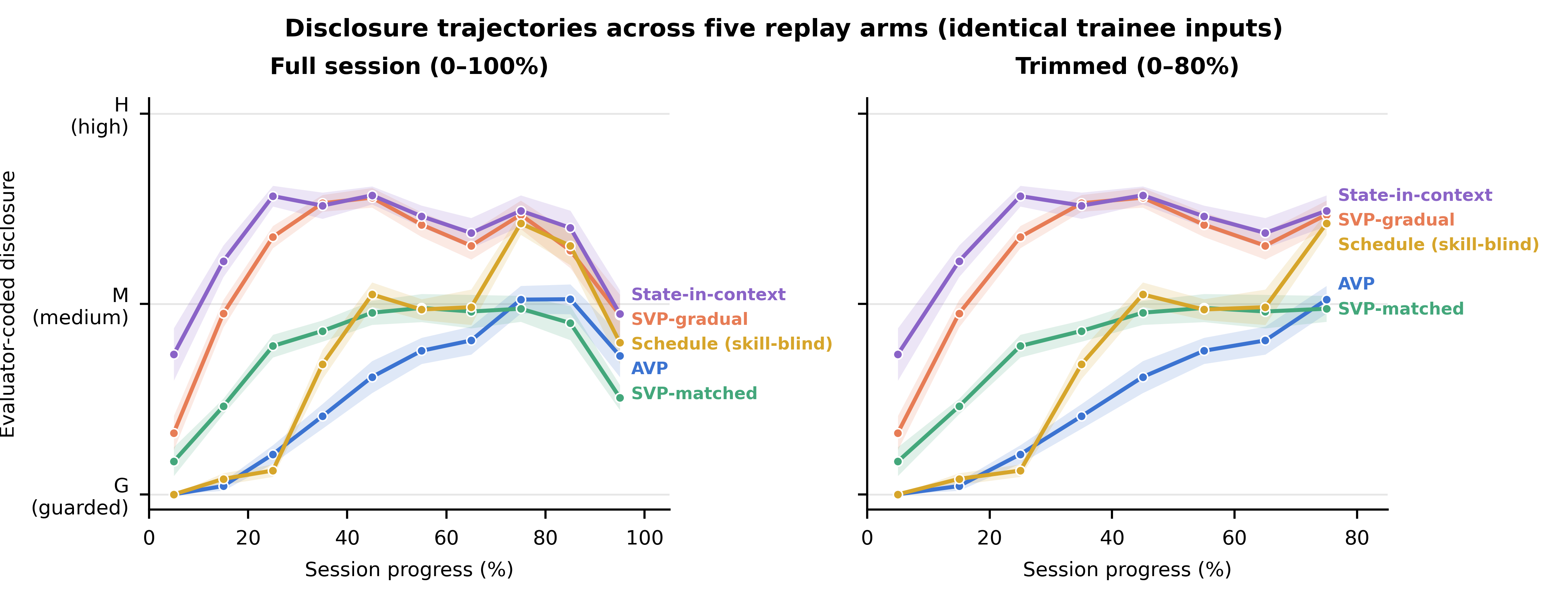}}%
    {\fbox{\parbox{0.95\textwidth}{\centering\vspace{2em}
      [Replay trajectory figure placeholder]\vspace{2em}}}}
  \caption{\textbf{Disclosure trajectories across 5 replay arms.}
{ 
    Evaluator-coded disclosure (G=1, M=2, H=3) is averaged by
    session-progress decile over identical replayed trainee turns.
    The left panel shows the full session (0--100\%); the right panel
    shows the active-interaction window (0--80\%).
    Bands indicate $\pm1$ SEM across participants after
    session-level aggregation.}}
  \label{fig:replay_trajectories}
\end{figure*}

\begin{table*}[t]
\centering\small
\caption{Human Likert ratings by condition (1--5 scale; $n = 20$ per condition). Means and standard deviations.}
\label{tab:human_ratings}
\begin{tabular}{lcccc}
\toprule
\textbf{Condition} & \textbf{Consistency} & \textbf{Adaptivity} & \textbf{Realism} & \textbf{Training Usefulness} \\
\midrule
Sam Alpha  (Static)   & 4.40 (0.68) & 3.20 (1.11) & 3.95 (0.76) & 3.95 (1.00) \\
Sam Beta  (Adaptive) & 4.55 (0.60) & 2.70 (0.92) & 4.05 (0.89) & 3.90 (0.97) \\
Alex Alpha (Static)   & 4.60 (0.60) & 3.75 (1.07) & 4.20 (0.77) & 4.05 (0.89) \\
Alex Beta (Adaptive) & 4.80 (0.41) & 4.00 (0.92) & 4.55 (0.76) & 4.25 (1.02) \\
\bottomrule
\end{tabular}
\end{table*}

\begin{table}[t]
\centering\small
\setlength{\tabcolsep}{4pt}
\caption{Ablation variants.}
\label{tab:ablation_variants}
\begin{tabular}{@{}lrrrr@{}}
\toprule
\textbf{Variant} & $w_i$ & $w_{er}$ & $w_r$ & $w_{ex}$ \\
\midrule
Deployed        & 1 & 1 & 1 & 3 \\
Uninformed      & 1 & 1 & 1 & 1 \\
Empathy-only    & 1 & 1 & 1 & 0 \\
Exploration-only & 0 & 0 & 0 & 3 \\
\bottomrule
\end{tabular}
\end{table}

\begin{table}[t]
\centering\small
\setlength{\tabcolsep}{5pt}
\caption{Evaluator-agreement under ablated weight schemes.
$\Delta$ = change in AVP agreement vs.\ Deployed.}
\label{tab:ablation_agreement}
\begin{tabular}{lrrrl}
\toprule
\textbf{Variant} & \textbf{AVP} & \textbf{SVP} & \textbf{Overall} & \textbf{$\Delta$} \\
\midrule
Deployed         & \textbf{.651} & \textbf{.311} & \textbf{.499} & --- \\
Uninformed       & .546          & .268          & .421          & $-$10.5 pp \\
Empathy-only     & .493          & .207          & .365          & $-$15.8 pp \\
Exploration-only & .579          & .238          & .426          & $-$7.2 pp \\
\bottomrule
\end{tabular}
\end{table}

\section{Secondary Analyses}\label{app:secondary}

This appendix collects two secondary observations moved out of the
main text.

\paragraph{Session length.} AVP sessions ran 2.7 turns longer than
SVP (14.3 vs.\ 11.6; $t(78)=3.83$, $p<.001$), a tentative engagement
signal. We report it primarily to rule out a confound: because
every trajectory analysis uses progress-normalized deciles, the
climbing-vs-flat contrast cannot be an artifact of differing session
lengths.

\paragraph{Model-based vs.\ human-rated adaptivity.} Pearson
correlations between the condition-level composite coupling score
and mean human ratings were weak across all dimensions
($r = .20$ for adaptivity, $r = -.39$ for realism, $r = -.23$ for
training usefulness). The automated coupling metric and
human-perceived responsiveness measure distinct aspects of system
behavior, motivating the property-based evaluation framework over
reliance on either alone.

\section{Role-Stratified Analysis}\label{app:role}

This appendix gives the per-dimension and head-to-head breakdown
of practitioner vs.\ student responses. The headline pattern is
that therapists rate every condition higher than students, with a
small gap for Sam and a large one for Alex (peaking at Alex Beta,
$\Delta=+0.61$). The gap is largest and most consistent on
perceived adaptivity (up to $+1.19$): therapists separate static
from adaptive Sam while students rate both at the floor. Crucially,
the \emph{objective} trajectory result is role-invariant: both
groups elicit a climbing AVP trajectory ($\Delta\approx+0.80$ to
$+0.95$) and a flat SVP trajectory, so the dominant signal is
system type, not user expertise.

\begin{table}[ht]
\centering\small\setlength{\tabcolsep}{5pt}
\caption{Mean overall rating by condition and role.}
\label{tab:role_overall}
\begin{tabular}{lrrl}
\toprule
\textbf{Condition} & \textbf{Stu.} & \textbf{Thr.} & \textbf{$\Delta$} \\
\midrule
Sam Alpha  (Static)   & 3.75 & 3.94 & $+0.19$ \\
Sam Beta   (Adaptive) & 3.75 & 3.83 & $+0.08$ \\
Alex Alpha (Static)   & 3.82 & 4.33 & $+0.51$ \\
Alex Beta  (Adaptive) & 4.00 & 4.61 & $+0.61$ \\
\bottomrule
\multicolumn{4}{l}{\scriptsize Stu.\ = Students ($n{=}7$);
  Thr.\ = Therapists ($n{=}13$). $\Delta$ = Thr.\ $-$ Stu.}
\end{tabular}
\end{table}

\begin{table}[ht]
\centering\small
\caption{Perceived adaptivity by condition and role.}
\label{tab:role_openness}
\begin{tabular}{lrrl}
\toprule
\textbf{Condition} & \textbf{Students} & \textbf{Therapists} & \textbf{$\Delta$} \\
\midrule
Sam Alpha  (Static)   & 2.43 & 3.62 & $+1.19$ \\
Sam Beta   (Adaptive) & 2.43 & 2.85 & $+0.42$ \\
Alex Alpha (Static)   & 3.29 & 4.00 & $+0.71$ \\
Alex Beta  (Adaptive) & 3.29 & 4.38 & $+1.09$ \\
\bottomrule
\multicolumn{4}{l}{\scriptsize Stu.\ = Students ($n{=}7$);
  Thr.\ = Therapists ($n{=}13$). $\Delta$ = Thr.\ $-$ Stu.}
\end{tabular}
\end{table}

\begin{table*}[ht]
\centering\small\setlength{\tabcolsep}{5pt}
\caption{Per-dimension ratings by condition and role
  (students $n=7$, therapists $n=13$).}
\label{tab:role_by_dim}
\begin{tabular}{lrrrrrrrr}
\toprule
& \multicolumn{2}{c}{\textbf{Consistency}}
& \multicolumn{2}{c}{\textbf{Adaptivity}}
& \multicolumn{2}{c}{\textbf{Realism}}
& \multicolumn{2}{c}{\textbf{Training}} \\
\cmidrule(lr){2-3}\cmidrule(lr){4-5}\cmidrule(lr){6-7}\cmidrule(lr){8-9}
\textbf{Condition} & \textit{Stu} & \textit{Thr}
  & \textit{Stu} & \textit{Thr}
  & \textit{Stu} & \textit{Thr}
  & \textit{Stu} & \textit{Thr} \\
\midrule
Sam Alpha  & 4.43 & 4.38 & 2.43 & 3.62 & 4.14 & 3.85 & 4.00 & 3.92 \\
Sam Beta   & 4.43 & 4.62 & 2.43 & 2.85 & 4.29 & 3.92 & 3.86 & 3.92 \\
Alex Alpha & 4.43 & 4.69 & 3.29 & 4.00 & 3.86 & 4.38 & 3.71 & 4.23 \\
Alex Beta  & 4.86 & 4.77 & 3.29 & 4.38 & 4.14 & 4.77 & 3.71 & 4.54 \\
\bottomrule
\end{tabular}
\end{table*}

\begin{table*}[ht]
\centering\small
\caption{Head-to-head preference votes (Alpha vs.\ Beta) by role.}
\label{tab:role_votes}
\begin{tabular}{llll}
\toprule
\textbf{Persona} & \textbf{Category} & \textbf{Students ($n=7$)} & \textbf{Therapists ($n=13$)} \\
\midrule
\multirow{4}{*}{Sam}  & Most responsive  & Alpha (5--2) & Alpha (10--3) \\
                      & More effective   & Beta  (4--3) & Beta  (7--6)  \\
                      & More realistic   & Alpha (4--3) & Beta  (7--6)  \\
                      & More challenging & Beta  (6--1) & Beta  (7--6)  \\
\midrule
\multirow{4}{*}{Alex} & Most responsive  & Alpha (4--3) & Alpha (7--6)  \\
                      & More effective   & Alpha (5--2) & Beta  (8--5)  \\
                      & More realistic   & Beta  (4--3) & Beta  (7--6)  \\
                      & More challenging & Beta  (6--1) & Beta  (12--1) \\
\bottomrule
\end{tabular}
\end{table*}

\section{Figures}\label{app:figures}

All figures in this appendix are referenced inline throughout the main paper.

\begin{figure*}[ht]
  \centering
  \IfFileExists{ui_screenshot.png}%
    {\includegraphics[width=\textwidth]{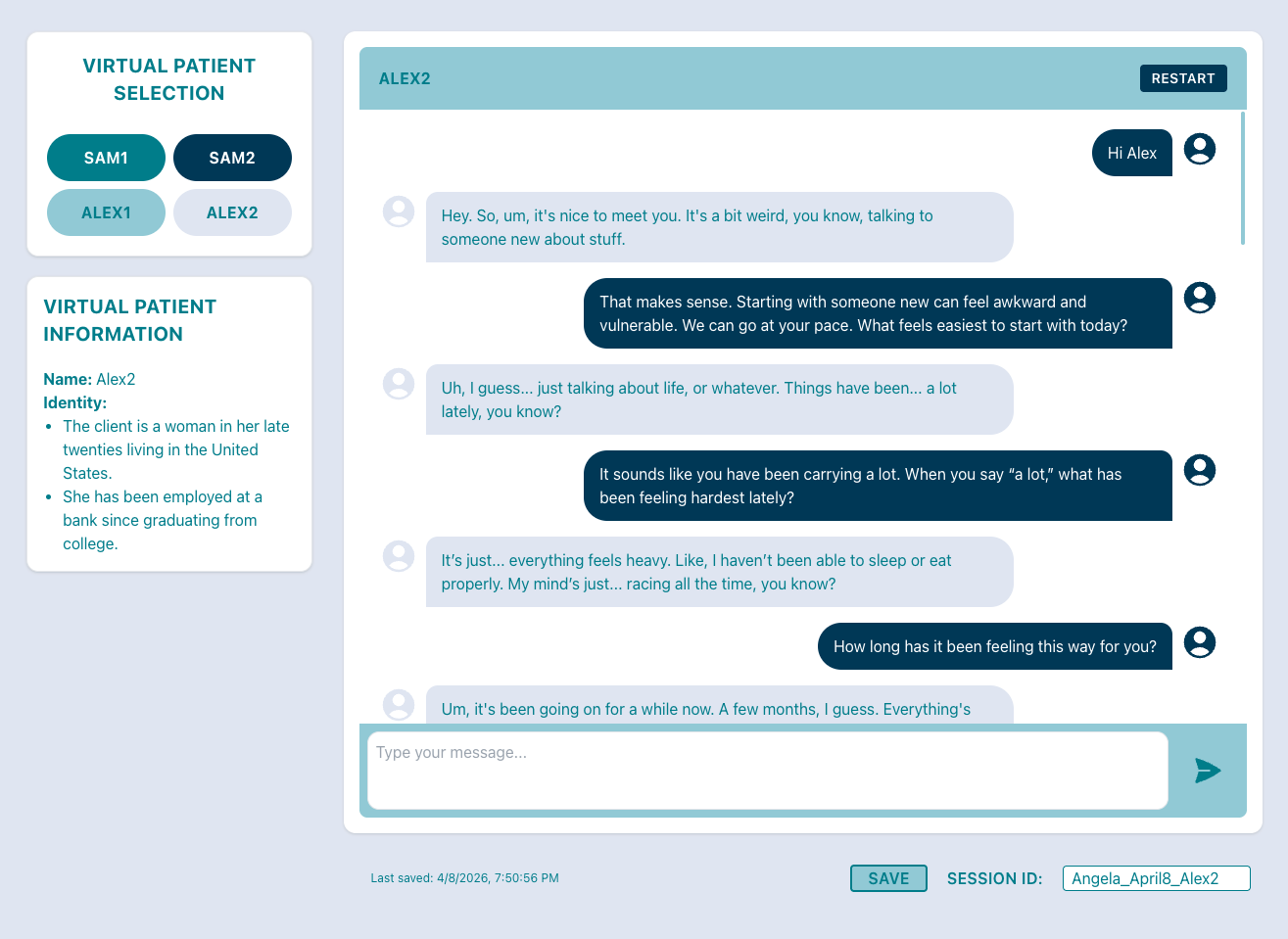}}%
    {\uifallback}
  \caption{\textbf{System interface.} Left panel: VP condition selection and
    patient background. Main panel: turn-by-turn chat workspace.
    A restart control allows the same case to be replayed from initial state.}
  \label{fig:ui}
\end{figure*}

\begin{figure}[ht]
  \centering
  \IfFileExists{fig_d3_trajectory.pdf}%
    {\includegraphics[width=\linewidth]{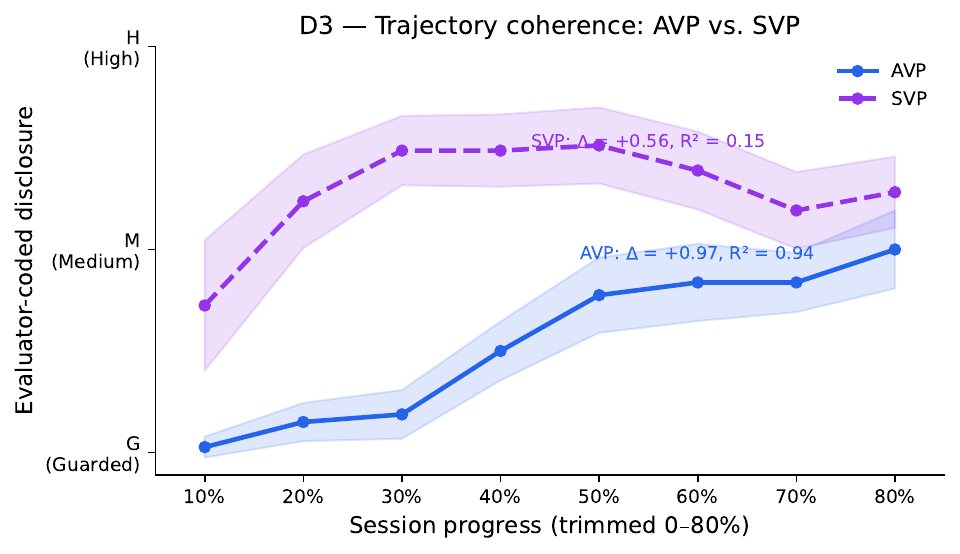}}%
    {\fbox{\parbox{0.95\linewidth}{\centering\vspace{2em}
      [Fig.\ A1 placeholder]\vspace{2em}}}}
  \caption{\textbf{P2 Gradual Build-up: AVP vs.\ SVP (trimmed 0--80\%).}
    Decile-averaged evaluator-coded disclosure (G=1, M=2, H=3) over session progress,
    trimmed to remove the closing-turn wrap-up artifact.
    AVP produces a strongly climbing trajectory ($\Delta = {+}0.97$, $R^2 = 0.94$);
    SVP is essentially flat ($\Delta = {+}0.56$, $R^2 = 0.15$).
    Shaded bands show 95\% CIs across sessions.
   }
  \label{fig:app_d3_trajectory}
\end{figure}

\begin{figure*}[ht]
  \centering
  \IfFileExists{fig_d3_persona.pdf}%
    {\includegraphics[width=\textwidth]{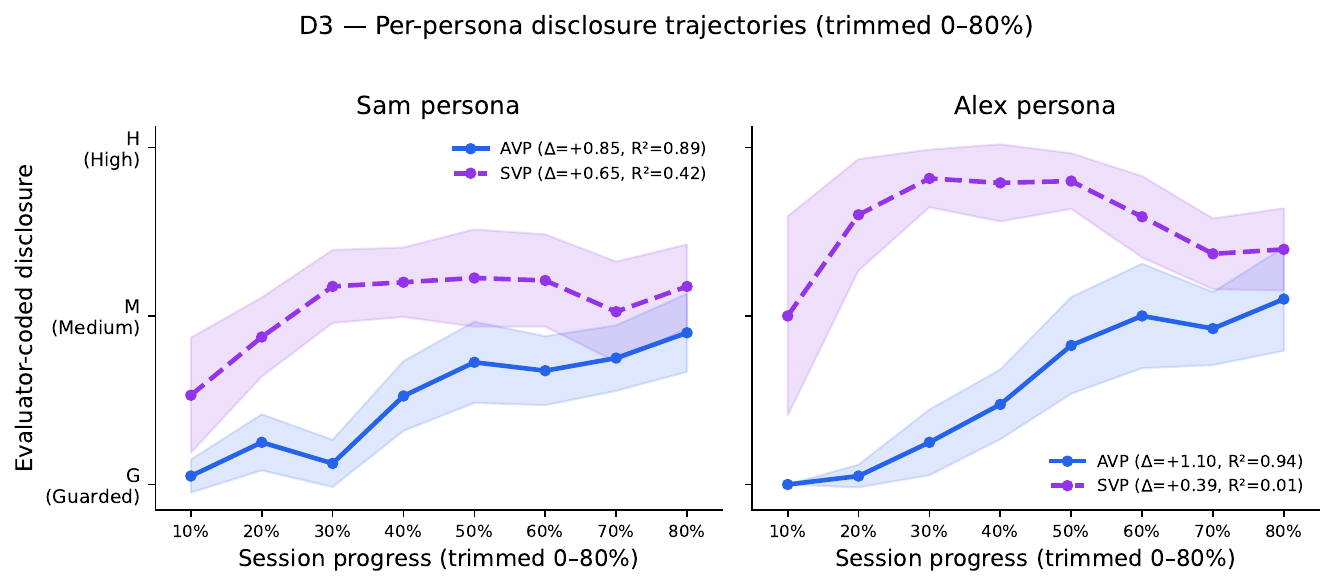}}%
    {\fbox{\parbox{0.95\textwidth}{\centering\vspace{2em}
      [Fig.\ A2 placeholder]\vspace{2em}}}}
  \caption{\textbf{P2 Per-persona disclosure trajectories (trimmed 0--80\%).}
    Left panel: Sam persona; right panel: Alex persona.
    The adaptive advantage is larger for Alex
    (AVP $\Delta = {+}1.10$, $R^2 = .94$; SVP $\Delta = {+}0.39$, $R^2 = .01$)
    than for Sam (AVP $\Delta = {+}0.85$, $R^2 = .89$; SVP $\Delta = {+}0.65$, $R^2 = .42$),
    consistent with Sam's adversarial persona eliciting a less distinct trajectory contrast.
    }
  \label{fig:app_d3_persona}
\end{figure*}

\begin{figure*}[ht]
  \centering
  \IfFileExists{fig_d1_coupling.pdf}%
    {\includegraphics[width=0.7\textwidth]{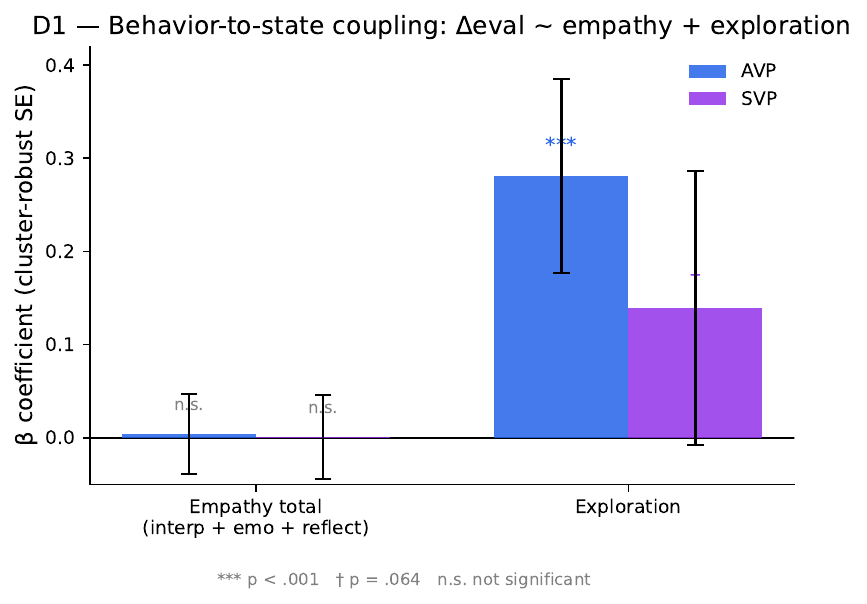}}%
    {\fbox{\parbox{0.65\textwidth}{\centering\vspace{2em}
      [Fig.\ A3 placeholder]\vspace{2em}}}}
  \caption{\textbf{P1 Responsiveness: $\hat\beta$ coefficients for empathy and
    exploration.}
    Grouped bars show OLS coefficients from
    $\Delta\text{eval} \sim \hat\beta_0 + \hat\beta_1\,\text{emp\_total} +
    \hat\beta_2\,\text{exploration}$ with cluster-robust SEs (session level).
    Exploration drives responsiveness in AVP ($\hat\beta = {+}0.281$, $p < 10^{-6}$);
    the empathy block is near zero in both designs ($\hat\beta < 0.005$, $p > .87$).
    Error bars = $\pm 1.96 \times$ SE.
    }
  \label{fig:app_d1_coupling}
\end{figure*}

\begin{figure*}[ht]
  \centering
  \IfFileExists{fig_d5_confusion.pdf}%
    {\includegraphics[width=\textwidth]{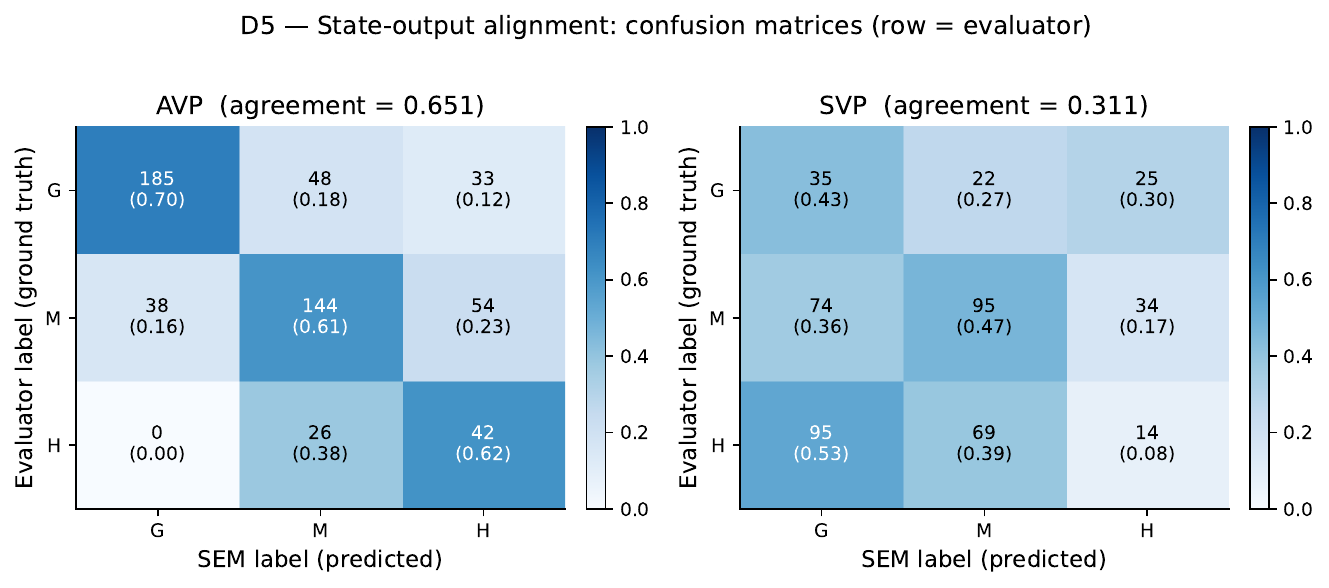}}%
    {\fbox{\parbox{0.95\textwidth}{\centering\vspace{2em}
      [Fig.\ A4 placeholder]\vspace{2em}}}}
  \caption{\textbf{P3 Faithful Realization (text alignment view): confusion matrices.}
    Cell values show counts and row-normalized proportions.
    AVP overall agreement = 0.651; G-precision = 0.83, indicating the state manager
    successfully restrains over-disclosure.
    SVP overall agreement = 0.311; SVP's H column is dominated by false positives
    (the always-H card causes the system to claim H-level disclosure while
    the evaluator observes G/M content), confirming therapist-invariant over-disclosure.
    }
  \label{fig:app_d5_confusion}
\end{figure*}

\begin{figure*}[ht]
  \centering
  \IfFileExists{fig_d6_discriminability.pdf}%
    {\includegraphics[width=\textwidth]{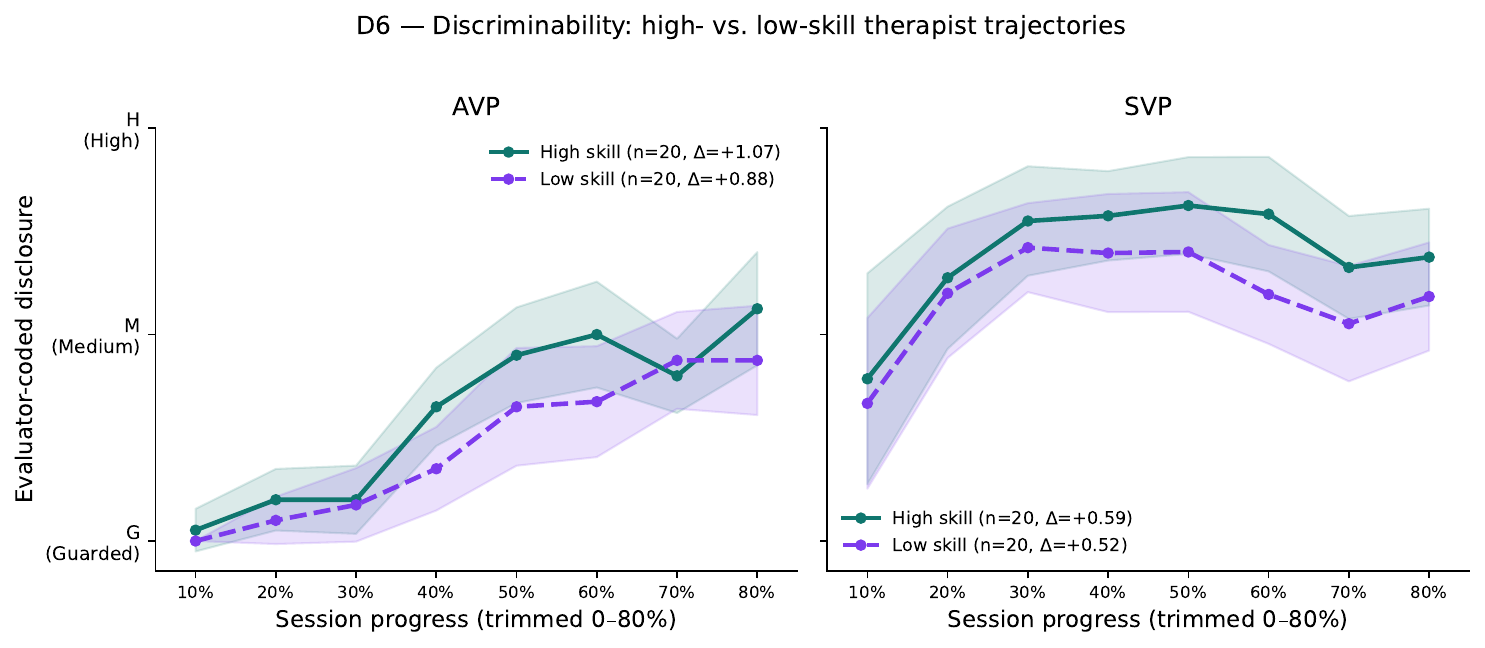}}%
    {\fbox{\parbox{0.95\textwidth}{\centering\vspace{2em}
      [Fig.\ A5 placeholder]\vspace{2em}}}}
  \caption{\textbf{P4 Faithful Realization (skill split view).}
    Sessions split at the median composite skill score
    ($\text{emp\_total} + 3{\times}\text{exploration}$, $n = 20$ per stratum).
    AVP shows a directional pattern: high-skill sessions ($\Delta = {+}0.85$) climb steeper
    than low-skill ($\Delta = {+}0.60$), though the formal slope $\times$ skill interaction
    is underpowered ($\hat\beta = {+}0.020$, $p = .27$).
    SVP trajectories are flat and skill-invariant.
    }
  \label{fig:app_d6_discriminability}
\end{figure*}

\begin{figure*}[ht]
  \centering
  \IfFileExists{fig_human_ratings.pdf}%
    {\includegraphics[width=\textwidth]{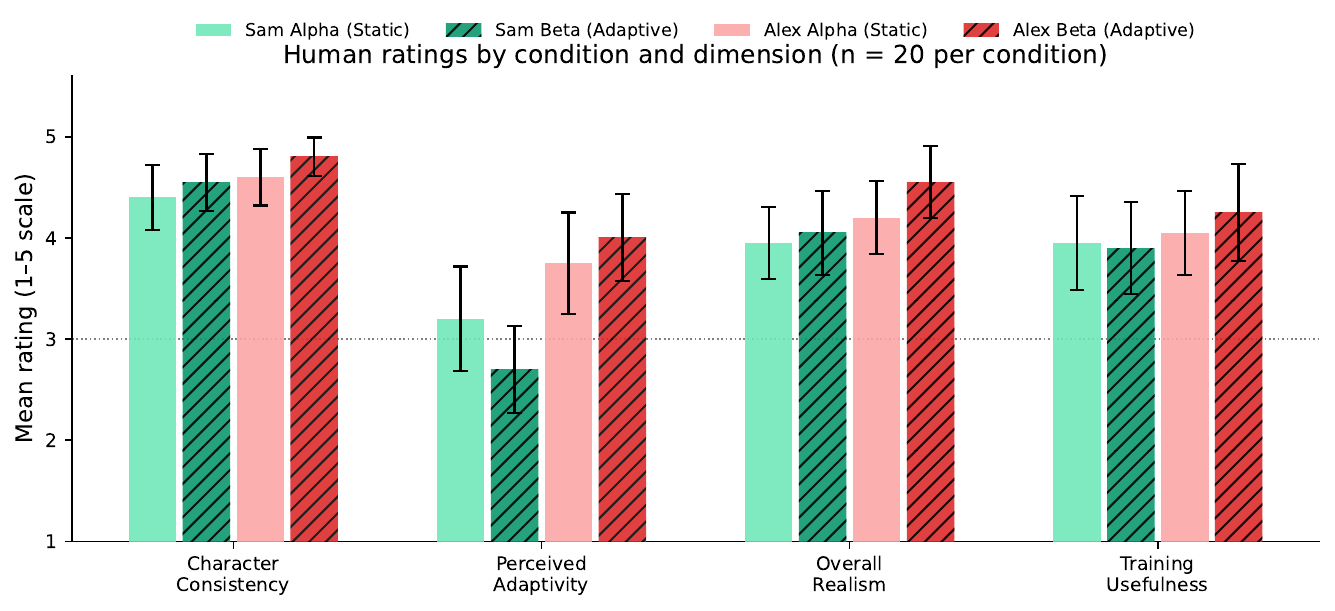}}%
    {\fbox{\parbox{0.95\textwidth}{\centering\vspace{2em}
      [Fig.\ A6 placeholder]\vspace{2em}}}}
  \caption{\textbf{Human Likert ratings by condition and dimension ($n = 20$ per condition).}
    Error bars show $\pm 1.96 \times$ SE.
    Solid fill = Static VP (Sam Alpha; Alex Alpha); hatched fill = Adaptive VP (Sam Beta; Alex Beta).
    Alex Beta (adaptive) is rated highest on all four dimensions; Sam Beta (adaptive) is rated
    \emph{lower} than Sam Alpha on perceived adaptivity, suggesting a persona-dependent
    reversal of the expected adaptivity advantage.
    }
  \label{fig:app_human_ratings}
\end{figure*}

\begin{figure*}[ht]
  \centering
  \IfFileExists{fig_ablation_agreement.pdf}%
    {\includegraphics[width=0.82\textwidth]{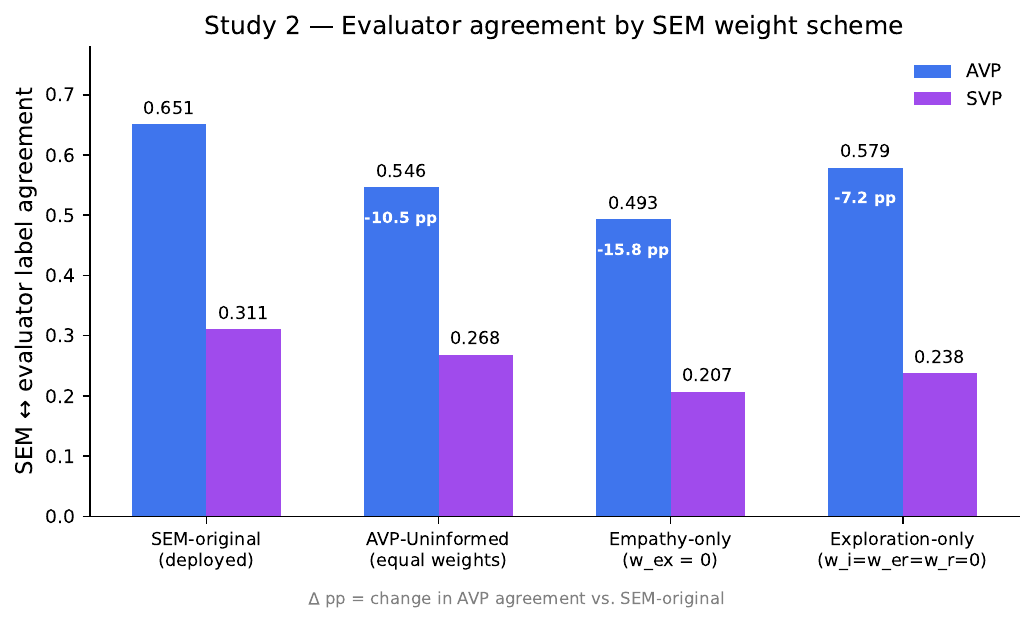}}%
    {\fbox{\parbox{0.78\textwidth}{\centering\vspace{2em}
      [Fig.\ A7 placeholder]\vspace{2em}}}}
  \caption{\textbf{Evaluator agreement by weight scheme (post-hoc ablation).}
    The deployed weighting outperforms all alternatives.
    Empathy-only loses 15.8 pp in AVP (the worst alternative) because without
    exploration the cumulative score cannot reach H-level disclosure.
    Exploration-only loses only 7.2 pp, confirming exploration as the dominant signal.
    }
  \label{fig:app_ablation_agreement}
\end{figure*}

\section{Survey Instrument}\label{app:survey}

This appendix reproduces the complete survey administered to participants.
The instrument comprised two parts: a \textbf{post-interaction survey} completed
after each of the four VP conditions, and a \textbf{final comparative survey}
completed after all four interactions were finished.

\subsection*{Post-Interaction Survey}

Participants completed this survey four times, once after each VP condition
(Sam Alpha, Sam Beta, Alex Alpha, Alex Beta).
\medskip

\noindent\textbf{Likert-scale items} \quad All four items used the same five-point scale:

\begin{center}
\small
\begin{tabularx}{\columnwidth}{@{}*{4}{>{\centering}X}>{\centering\arraybackslash}X@{}}
\textbf{1} & \textbf{2} & \textbf{3} & \textbf{4} & \textbf{5} \\[3pt]
\textit{Very slightly or not at all} &
\textit{A little} &
\textit{Moderately} &
\textit{Quite a bit} &
\textit{Extremely}
\end{tabularx}
\end{center}

\begin{enumerate}[label=\textbf{Q\arabic*.}, leftmargin=*, itemsep=4pt]
  \item \textbf{Character consistency.}
        To what extent did the virtual patient's responses stay in their character? Please rate and briefly explain your reasoning.
  \item \textbf{Perceived adaptivity.}
        To what extent did the virtual patient gradually shift its openness in response to your empathetic engagement? Please rate and briefly explain your reasoning.
  \item \textbf{Overall realism.}
        To what extent did the virtual patient feel realistic? Please rate and briefly explain your reasoning.
  \item \textbf{Training usefulness for empathy practice.}
        Based on the conversation content alone, to what extent do you think this system helps a trainee practice the following empathy skills: exploration, interpretation, reflective listening, and emotional reaction? Please rate and briefly explain your reasoning.
\end{enumerate}

\subsection*{Final Comparative Survey}

After completing all four interactions, participants received the following
framing prompt before answering the structured questions below.
\medskip

\noindent\textbf{SAM comparison questions} \quad
Asked about SAM Alpha (static VP) vs.\ SAM Beta (adaptive VP).

\begin{enumerate}[label=\textbf{S\arabic*.}, leftmargin=*, itemsep=4pt]
  \item Which virtual patient (SAM Alpha, SAM Beta) felt more
        \textbf{responsive or adaptive} to how you were engaging with it?
        What made that one stand out? \hfill[\textit{choice + open text}]
  \item Which felt more \textbf{effective for simulating a therapeutic
        interaction}, and why? \hfill[\textit{choice + open text}]
  \item Which felt more \textbf{realistic} to you, and why?
        \hfill[\textit{choice + open text}]
  \item Did one feel more \textbf{challenging} to engage with?
        What contributed to that impression? \hfill[\textit{choice + open text}]
  \item How would you describe the \textbf{behavioral differences} between
        SAM Alpha and SAM Beta? \hfill[\textit{open text}]
\end{enumerate}

\noindent\textbf{ALEX comparison questions} \quad
The same five questions were repeated for ALEX Alpha vs.\ ALEX Beta.

\medskip
\noindent\textbf{General reflective questions}

\begin{enumerate}[label=\textbf{G\arabic*.}, leftmargin=*, itemsep=4pt]
  \item Compared to any \textbf{traditional training approaches} you have
        experienced, how did your experience with the virtual patient differ?\hfill[\textit{open text}]
  \item Do you have any \textbf{suggestions for improving the realism and
        training usefulness} of the virtual patient?
        \hfill[\textit{open text}]
\end{enumerate}

\section{Participant Procedures and Ethics}\label{app:procedure}

This appendix documents participant instructions, recruitment and compensation, informed consent procedures, and ethics review.

\subsection*{Instructions Given to Participants}
Participants received a survey with a written study brief by email prior to the session and an oral walkthrough at session start. The brief described the study as an evaluation of LLM-based virtual patients for psychotherapy training. The oral walkthrough explained that they would interact with four virtual patients and stated that they would not be evaluated by their clinical skills. Participants were told that (i) the order of personas and of system conditions within each persona pair was randomized; and (ii) each interaction would be followed by a short post-interaction survey, with a final comparative survey after all four interactions.

Risks of participation were disclosed to participants in the Risks section of the consent form, which read in full:
\textit{The risks and discomfort associated with participation in this study are no greater than those ordinarily encountered in psychotherapy training. There is a risk of potentially private or sensitive information being discussed during the conversations. There is risk of potential breach of confidentiality if the transcripts of conversations with the virtual patients are inadvertently released.}
\subsection*{Recruitment and Payment}
Participants were recruited through targeted email invitations sent to licensed clinicians in the authors' professional networks and clinical psychology students from different institutions. The invitation described the study purpose, eligibility, expected time commitment, compensation, and contact information for questions. No crowdsourcing platforms were used; all participants were identifiable professionals or trainees recruited directly. Of the 20 participants, 13 were licensed practitioners and 7 were clinical students with relevant psychology training. All participants were located in the United States. Therapist participants received \$225 (\$150/hour), and student participants received \$30 (\$20/hour) for completing the full 90-minute session.

\subsection*{Data Consent}
Before any interaction, each participant reviewed the online consent form and indicated consent by affirmatively responding to three items confirming that they were 18 years or older, had read and understood the study information, and wished to participate. No physical signature was collected; affirmative responses to all three items served as consent. The consent form described: (i) the study purpose, namely investigating methods to improve a listener-training chatbot through interaction with realistic simulated members; (ii) study procedures, including reviewing scenarios, interacting with prototype virtual patients, providing feedback, and completing a short follow-up interview, all conducted online and lasting no more than 90 minutes; (iii) collected data, including interaction transcripts, survey and interview responses, role information (e.g., practitioner or student status), and video/audio recordings of the session; (iv) confidentiality protections, including separate storage of consent forms and study data on secure institutional servers accessible only to project investigators, with research records retained for at least three years; (v) participants' right to withdraw at any time; and (vi) future use of de-identified data for future research or sharing with other researchers without additional consent. Participants were informed that video recording could be stopped at any time and that any portion of the recording could be erased upon request.
\subsection*{Ethics Review Board Approval}
The data collection protocol was reviewed and approved by the Institutional Review Board (IRB) at the authors' institution.  

\end{document}